\documentclass[lettersize,journal]{IEEEtran}
\usepackage{amsmath,amsfonts}
\usepackage{algorithmicx}
\usepackage{algorithm}
\usepackage{array}
\usepackage[caption=false,font=footnotesize,labelfont=rm,textfont=rm]{subfig}
\usepackage{textcomp}
\usepackage{stfloats}
\usepackage{url}
\usepackage{verbatim}
\usepackage{graphicx}
\usepackage{cite}
\usepackage{soul, color, xcolor}
\usepackage[mathscr]{euscript}
\usepackage{booktabs}
\usepackage{algpseudocode}
\usepackage{hyperref}
\usepackage{orcidlink}
\usepackage{float}  
\begin{document}

\title{Cooperative Sensing Enhanced UAV Path-Following and Obstacle Avoidance with Variable Formation}

\author{Changheng~Wang, Zhiqing~Wei,~\IEEEmembership{Member,~IEEE,} Wangjun~Jiang,~\IEEEmembership{Student Member,~IEEE,} \\Haoyue~Jiang, and Zhiyong~Feng,~\IEEEmembership{Senior Member,~IEEE}
\thanks{
Changheng Wang, Zhiqing Wei, Wangjun Jiang, Haoyue Jiang, and Zhiyong Feng are with Beijing University of Posts and Telecommunications, Beijing 100876, China (e-mail: ch\_wang@bupt.edu.cn, weizhiqing@bupt.edu.cn, jiangwangjun@bupt.edu.cn, jiang\_hy120@bupt.edu.cn, fengzy@bupt.edu.cn). Correspondence author: Zhiqing Wei.}}


\maketitle

\begin{abstract}
The high mobility of unmanned aerial vehicles (UAVs) enables them to be used in various civilian fields, such as rescue and cargo transport. Path-following is a crucial way to perform these tasks while sensing and collision avoidance are essential for safe flight. In this paper, we investigate how to efficiently and accurately achieve path-following, obstacle sensing and avoidance subtasks, as well as their conflict-free fusion scheduling. Firstly, a high precision deep reinforcement learning (DRL)-based UAV formation path-following model is developed, and the reward function with adaptive weights is designed from the perspective of distance and velocity errors. Then, we use integrated sensing and communication (ISAC) signals to detect the obstacle and derive the Cramér-Rao lower bound (CRLB) for obstacle sensing by information-level fusion, based on which we propose the variable formation enhanced obstacle position estimation (VFEO) algorithm. In addition, an online obstacle avoidance scheme without pretraining is designed to solve the sparse reward. Finally, with the aid of null space based (NSB) behavioral method, we present a hierarchical subtasks fusion strategy. Simulation results demonstrate the effectiveness and superiority of the subtask algorithms and the hierarchical fusion strategy.
\end{abstract}


\section{Introduction}
\IEEEPARstart{W}{ith} the development of wireless communication and intelligent control, unmanned aerial vehicle (UAV) plays an important role in the civilian field. However, the detection and communication capabilities of a single UAV are limited. Multiple UAVs provide enhanced coverage and stability \cite{khan2022cooperative}, enabling them to perform various tasks such as rescue, cargo transport, emergency communications, and area search \cite{55JavaidCommunication}. Path-following serves as a prerequisite for completing these tasks.

One of the important ways for multi-UAVs to perform tasks is UAV formation cooperative path-following (PF) \cite{sen2020information}, which faces several challenges. Firstly, UAVs equipped with both communication and radar equipment have large hardware loads and low spectral and energy efficiency. In addition, a simple and efficient path-following algorithm is a prerequisite for practical availability. Furthermore, UAV formation requires the ability to sense the obstacle accurately. Finally, cooperative obstacle avoidance (OA) is critical to ensuring flight safety. Therefore, this paper focuses on efficient and accurate algorithms for path-following, obstacle sensing, and obstacle avoidance subtasks, as well as a fusion strategy for these subtasks.

For the UAV formation path-following, there have been many excellent works. Chen et al. \cite{chen2021coordinated} proposed to handle the forward and angular velocity constraints of fixed-wing UAVs through a hybrid control law, guiding UAV formation to fly along the path. In \cite{cao2022concentrated}, consensus theory was used to guide the UAV formation to follow the coverage path in combination with the artificial potential field method. With the development of artificial intelligence technology, deep reinforcement learning (DRL) was used to solve the UAV path-following problem. In \cite{xia2021multi}, the soft actor-critic (SAC) algorithm was used to track the moving target based on location information, introducing spatial information entropy (SIE) to maintain the distance between UAVs. In \cite{zhao2021usv}, a path-following scheme based on deep deterministic policy gradient (DDPG) was proposed for unmanned surface vessel (USV) formation, and an efficient reward function was designed from the perspective of formation efficiency and distance error. A two-stage reward function for fixed-wing UAV formation path-following was proposed in \cite{shi2022leader}, related to the penalty of boundary and formation error. In \cite{wei20223u}, the cooperative underwater target enclosing network was designed with separate rewards for encircling and tracking constraints. It is worth noting that the reward functions in the majority of DRL path-following algorithms were designed with the distance error, without considering the velocity error with the target. To enhance accuracy, it is essential to design the reward function considering both distance and velocity errors. Furthermore, the above DRL-based methods, which are relatively complex, were dedicated to training for the entire UAV formation. However, the trained model may fail when the number of UAVs changes.

In terms of UAV formation obstacle sensing, including both vision-based and nonvisual detection methods \cite{wei2021anti}, where radar detection in nonvisual can take advantage of integrated sensing and communication (ISAC) and avoid the sensitivity to light intensity in visual detections. Moreover, the precision of obstacle estimation is greatly related to the number of UAVs and the formation shape \cite{chand2017sense}, which directly affects the obstacle avoidance performance. Wang et al. \cite{51Feasibility} proposed a ground user positioning method based on time difference of arrival (TDOA) and double-response two-way ranging. In \cite{52TDOAAOA}, a non-line-of-sight (NLoS) localization method based on TDOA and angle of arrival (AOA) was proposed to achieve localization by splitting into two line-of-sight (LoS) paths modeled separately. In \cite{53UAV}, an extended Kalman filter (EKF) with a colored noise model was proposed to improve the accuracy of the received signal strength (RSS)-based localization algorithm. However, the above methods do not take into account the improvement of target sensing performance through changes in sensor positions. The localization of an unknown moving target by heterogeneous UAV-aware networks was studied in \cite{sen2020information} to maximize target state information by generating desired formation, but it does not consider the case that the UAVs do not surround the target. Meng et al. \cite{meng2016optimal} studied the optimal geometry of TDOA-based multiple UAV sensors for mobile source localization, but the formation is loosely and may not be suitable for general path-following tasks.

In UAV formation obstacle avoidance, classical geometric methods \cite{tan2020three}, graph theory methods \cite{zhang2018quantitative}, artificial potential field \cite{hu2020formation}, and path selection-based methods \cite{ahmed2021energy} have matured relatively. Machine learning (ML) is also gradually being applied in this domain. Wang et al.\cite{wang2021collision} proposed a UAV trajectory design using dueling double deep Q-network incorporating collision avoidance and communication constraints. In \cite{xue2022uav}, a fast recursive random valued gradient algorithm based on the actor-critic framework was designed to solve the safe navigation of UAVs in unknown environments containing numerous obstacles. To solve the dynamic obstacle avoidance problem, \cite{han2019multi} proposed the experience-shared advantaged actor-critic (A2C) algorithm in which the UAV can share the learned experience with other UAVs in the cluster. In \cite{wang2021collision,xue2022uav,han2019multi}, collision avoidance is integrated into the reward function in conjunction with path-following, significantly increasing the complexity of model training. In practice, spatial obstacles appear randomly, and only sparse rewards are available, requiring extensive training in the model. The combination of DRL-based path-following and classical obstacle avoidance schemes can effectively solve the sparse reward problem, and avoid retraining the following path after the obstacle avoidance is completed \cite{wei2021anti}.

To efficiently and accurately complete the three subtasks of path-following, obstacle sensing, and avoidance, there are high-performance requirements for both communication and sensing. As a radio communication technology with ultra-high data rate, ultra-low latency, and ultra-high density coverage, the 5th-generation mobile communication technology (5G) can provide reliable communication for UAV formation \cite{li2018uav}. Meanwhile, the pilot signals in 5G can be used for radar detection \cite{40wei20225g}, enabling ISAC. On the one hand, ISAC technology allows UAVs to simultaneously carry radar sensing and communication functions under load, volume, spectrum, and power constraints, reducing hardware costs while improving spectrum and energy efficiency \cite{54JRCsurvey}. On the other hand, the joint processing of ISAC information to exploit the sensing potential of communication signals realizes fast and accurate target identification and localization in most scenarios \cite{meng2023uav}. In \cite{jiang2021improve}, the EKF was used to fuse communication position information with sensing information obtained from ISAC signals, improving target sensing accuracy. In \cite{yang2023novel}, a high precision ISAC geometry-based stochastic model was designed, using the position estimates of the first and last bounce scatterers in the communication channel. To improve spectral efficiency, DRL was used in \cite{qin2023deep} to address joint user association, UAV trajectory planning, and power allocation problem. In \cite{chen2020performance}, a novel ISAC antenna consisting of sensing and communication subarrays was proposed to achieve beam sharing and superior sensing performance. Moreover, the LoS links of traditional terrestrial ISAC networks are susceptible to blocking by obstacles \cite{meng2023uav}. In contrast, UAV networks have fewer obstacles and high probability of LoS links, which can better utilize the advantages of ISAC.

In this paper, ISAC technology is used to improve the sensing performance of UAV formation, reduce hardware costs and communication overhead. To tackle several challenges in UAV formation path-following tasks, we propose a novel strategy. This strategy aims to improve accuracy and scalability of path-following, address imprecisions in spatial obstacle sensing, and overcome the challenge of achieving convergence with sparse rewards in learning-based obstacle avoidance algorithms. It is a cooperative sensing enhanced UAV formation path-following and obstacle avoidance strategy with variable formation based on ISAC signal, where the following target is virtual-leader (VL). In the process of path-following, the transformation of UAV formation can improve the accuracy of obstacle sensing, and the precise obstacle sensing can conversely enhance the success rate of obstacle avoidance, thereby facilitating mutual promotion of cooperative obstacle sensing and UAV formation control performance. The formation consists of a master UAV (MUAV) and several auxiliary UAVs (AUAVs). The MUAV performs information-level fusion of ISAC sensing information to obtain obstacle position and velocity estimates, and issues virtual following targets (VFTs). Each AUAV makes independent control decisions based on its position and the received VFT from MUAV. The main contributions of this paper are summarized as follows.

\begin{enumerate}
	{\item We propose a DRL-based UAV formation path-following algorithm and design a reward function with adaptive weights, considering distance and velocity errors. The weight for velocity error increases as distance error decrease, enabling high-precision path-following without added complexity or compromised convergence speed.}
	{\item We propose a formation-variable dynamic obstacle sensing algorithm, and drive the Cramér-Rao lower bound (CRLB) for obstacle sensing based on ISAC signal. With the aim of minimizing the CRLB, we find the optimal UAV positions to enhance the precision of obstacle position estimation.}
	{\item Based on the DRL path-following model, the online obstacle avoidance scheme is designed to solve the sparse reward problem without pretraining.}
	{\item We propose a null space based hierarchical subtasks fusion strategy. To ensure the safe and conflict-free flight of UAVs, sensing and obstacle avoidance are assigned the highest priority, while path-following is projected into the null space of obstacle avoidance.} Finally, the simulation results verify the effectiveness and superiority of the proposed algorithms.
\end{enumerate}

The rest of the paper is organized as follows. Section II introduces the UAV kinetics, UAV formation model, and obstacle dynamics. Section III presents the information-level fusion scheme for the obstacle position and velocity estimates, and derives its CRLB. Section IV introduces the proposed DRL path-following algorithm, variable formation enhanced obstacle position estimation algorithm and their complexity analysis, as well as the online obstacle avoidance algorithm. Section V proposes a hierarchical subtasks fusion strategy. The simulation results are presented in Section VI. The conclusions are drawn in Section VII.

\section{System Model}

\begin{figure}[!t]
	\centering
	\includegraphics[width=3.5in]{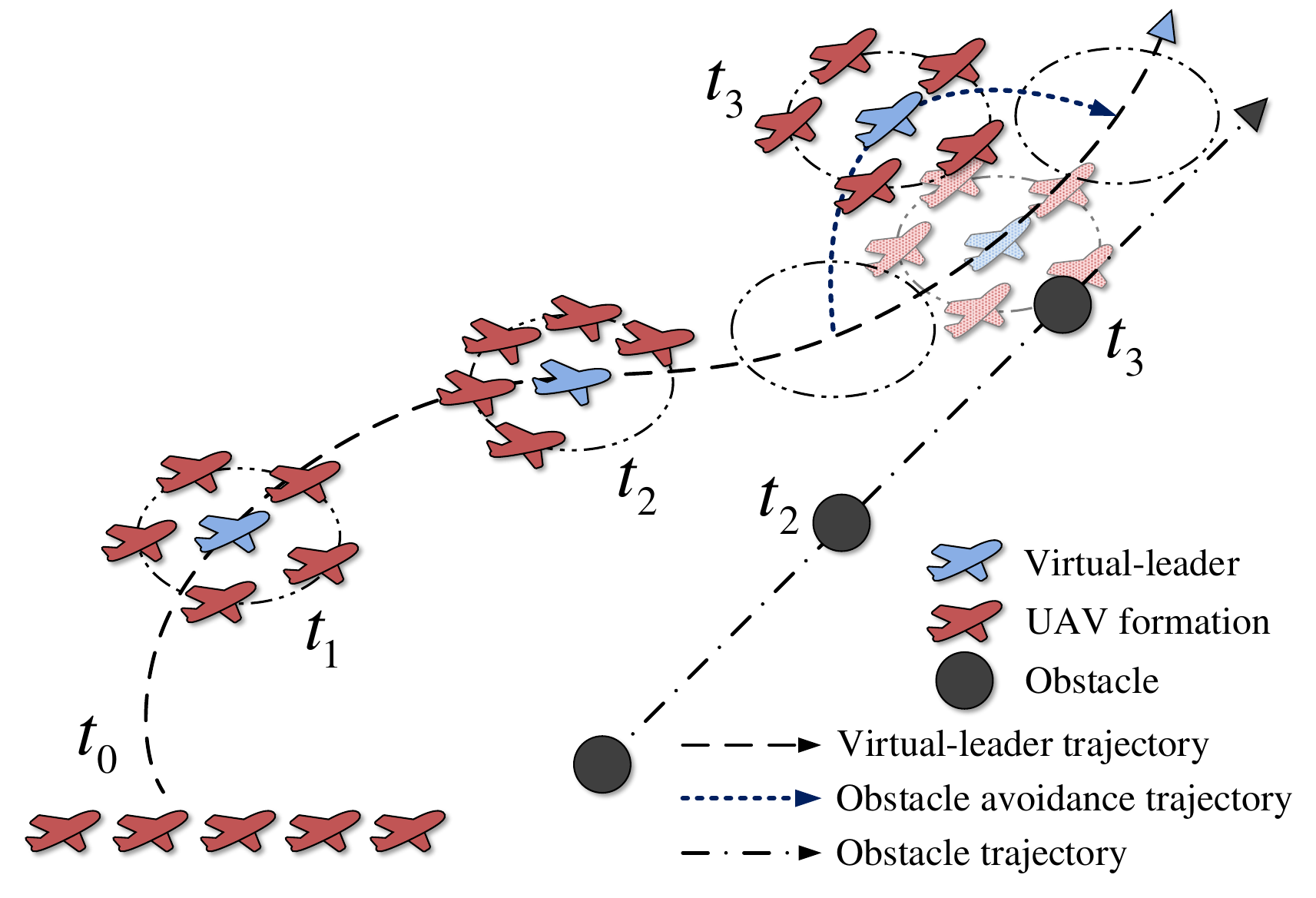}
	\caption{Tasks scenario. $t_0$: UAVs initial time, $t_1$: UAV formation path-following time, $t_2$: Variable formation enhances obstacle sensing time, $t_3$: Obstacle avoidance time.}
	\label{fig_1}
\end{figure}

In this paper, we consider the UAV formation obstacle avoidance scenario in three-dimensional (3D) space. As shown in Fig. \ref{fig_1}, the UAV formation is centered on the virtual-leader to follow a predetermined trajectory, and avoid collision with the dynamic obstacle. The virtual-leader is a virtual reference point set to facilitate UAV formation flight.

During the process of the UAV formation performing virtual-leader path-following tasks, three subtasks are involved in path-following, obstacle sensing, and obstacle avoidance. As shown in Fig. \ref{fig_1}, multi-UAVs take off to chase the virtual-leader in the first phase ($t_0$). In the second phase ($t_1$), the UAVs have caught up with the virtual-leader and formed a predetermined formation to trigger path-following. In the third phase ($t_2$), UAV formation senses the obstacle and improves the accuracy of the obstacle observation by changing the position of UAVs in the formation. In the fourth phase ($t_3$), the distance between the obstacle and the UAV formation is less than the minimum safety distance. The obstacle avoidance subtask is initiated and continues to follow the virtual-leader path once the obstacle avoidance is completed.

\subsection{UAV Kinetics}

Consider there exists $P$ UAVs in the formation, and each UAV has the same mechanical structure. The kinetic model of the $i$-th UAV (UAV $i$) is \cite{zhang2021leader}
\begin{equation}\label{1}
	\begin{matrix}  
 	    \dot{x}_i=V_{i}\cos{\gamma_i}\cos{\chi_i},
    \end{matrix}
\end{equation}
\begin{equation}\label{2}
	\begin{matrix}  
		\dot{y}_i=V_{i}\cos{\gamma_i}\sin{\chi_i},
	\end{matrix}
\end{equation}
\begin{equation}\label{3}
	\begin{matrix}  
		\dot{z}_i=V_{i}\sin{\gamma_i},
	\end{matrix}
\end{equation}
\begin{equation}\label{4}
	\begin{matrix}  
		\dot{V}_i=\frac{T_i-D_i}{m_i}-g\sin{\gamma_i},
	\end{matrix}
\end{equation}
\begin{equation}\label{5}
	\begin{matrix}  
		\dot{\gamma}_i=\frac{g}{V_i}(\frac{L_i-\cos{\phi_i}}{g m_i}-\cos\gamma_i),
	\end{matrix}
\end{equation}
\begin{equation}\label{6}
	\begin{matrix}  
		\dot{\chi}_i=\frac{L_i\sin\phi_i}{m_iV_i\cos\gamma_i},
	\end{matrix}
\end{equation}

\begin{figure}[!t]
	\centering
	\includegraphics[width=2.5in]{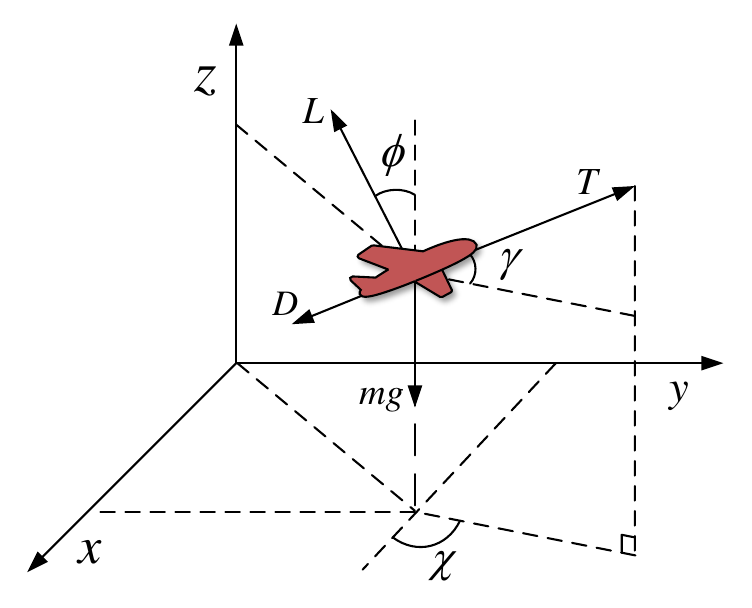}
	\caption{UAV kinetics in the inertial coordinate system.}
	\label{fig_2}
\end{figure}

\noindent where $i=1,2,\ldots,P$. Establish an inertial coordinate system as shown in Fig. \ref{fig_2}, where the $x$-axis, $y$-axis, and $z$-axis point to the east, north, and vertically upward from the center of the earth respectively. $V_i$ is ground speed, assumed to be equal to airspeed. $\dot{x}_i$, $\dot{y}_i$ and $\dot{z}_i$ are the velocity components of $V_i$ in the direction of the east-north-up coordinate axes. $m_i$ is the UAV $i$ mass, $g$ is the acceleration of gravity. $\chi_i$, $\gamma_i$ are the track angle and heading angle. $D_i$, $L_i$ are the drag and vehicle lift respectively.

In the UAV kinetic model, the load factor $n_i$ controlled by the elevator, the banking angle $\phi_i$ controlled by the combination of the rudder and the aileron, and the engine thrust $T_i$ controlled by the throttle are the actual control variables. \eqref{1}, \eqref{2}, \eqref{3} are differentiated once for time and then substituted into the actual UAV control variables as follows \cite{zhang2021leader}.

\begin{small}
\begin{equation} \label{7}
		\phi_i=\tan^{-1}{\Big[\displaystyle\frac{a_{y,i}\cos\chi_i-a_{x,i}\sin\chi_i}{\cos\gamma_i(a_{z,i}+g)-\sin\gamma_i(a_{x,i}\cos\chi_i+a_{y,i}\sin\chi_i)}\Big]},
\end{equation}
\begin{equation} \label{8}
		n_i=\displaystyle\frac{\cos\gamma_i(a_{z,i}+g)-\sin\gamma_i(a_{x,i}\cos\chi_i+a_{y,i}\sin\chi_i)}{g\cos\phi_i},
\end{equation}
\begin{equation}\label{9}
		T_i=[\sin\gamma_i(a_{z,i}+g)+\cos\gamma_i(a_{x,i}\cos\chi_i+a_{y,i}\sin\chi_i)]m_i+D_i,
\end{equation}
\end{small}

\noindent where $a_{x,i}$, $a_{y,i}$ and $a_{z,i}$ are the acceleration of UAV $i$ in the directions of the three coordinate axes. $\chi_i=\arctan(\dot{y}_i/\dot{x}_i)$, $\gamma_i=\arctan(\dot{z}_i/V_i)$. In view of this, the new virtual control variables $\boldsymbol{\alpha}_i=[a_{x,i},a_{y,i},a_{z,i}]^\mathrm{T}$ are obtained, which can be inserted into \eqref{7}, \eqref{8}, \eqref{9} to obtain the real control input.

\subsection{UAV Formation}

\begin{figure}[!t]
	\centering
	\includegraphics[width=2.5in]{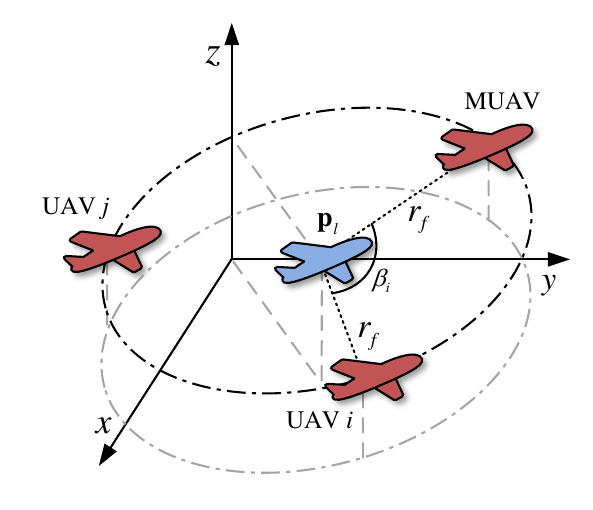}
	\caption{UAV formation model.}
	\label{fig_3}
\end{figure}

The virtual-leader is introduced to calibrate the reference position of each UAV path-following in the formation, and the virtual-leader is located at the center of the formation. As shown in Fig. \ref{fig_3}, the position of virtual-leader is $\boldsymbol{p}_l=[x_l,y_l,z_l]^\mathrm{T}$. The UAV formation includes one MUAV and $P-1$ AUAVs. During the performance of the tasks, the MUAV calculates the VFT of each UAV according to the number of UAVs in the formation and the position of the virtual-leader, and sends to AUAVs, as UAV $i$ and UAV $j$. In this way, the formation is formed and the path-following task is carried out. The VFT $\boldsymbol{u}_{ie}=[x_{ie},y_{ie},z_{ie}]^\mathrm{T}$ of the UAV $i$ in the formation is
\begin{equation}\label{10}
		x_{ie}=x_l+r_f\cos\Big(\arctan\big(\frac{y_l}{x_l}\big)+\beta_i\Big),
\end{equation}
\begin{equation}\label{11}
		y_{ie}=y_l+r_f\sin\Big(\arctan\big(\frac{y_l}{x_l}\big)+\beta_i\Big),
\end{equation}
\begin{equation}\label{12}
		z_{ie}=z_l.
\end{equation}

In the process of path-following, all UAVs in the formation lie on the virtual circle with center $\boldsymbol{p}_l$ and radius $r_f$. However, the UAV formation is not always constant. Usually, UAVs in the formation are evenly distributed on the virtual circle with the same angle $\beta_i$ at the equal distance $r_f$ from the virtual-leader. Once a uniformly distributed UAV formation cannot meet the position estimation requirement of the spatial obstacle, namely, when the CRLB of the UAV formation for obstacle observation is greater than the threshold, the CRLB is minimized by altering the position of UAVs in the formation to obtain a more accurate obstacle position estimation. During the UAV formation transform, the UAVs persist in their arrangement along the virtual circle at an equidistant $r_f$ from the virtual-leader. However, it is noteworthy that this distribution does not adhere to uniformity. At this time, the VFTs of UAVs in the formation are also allocated by MUAV.

At time $k$, the error between the actual position $\boldsymbol{u}_i(k)=[x_i,y_i,z_i]_k^\mathrm{T}$ and the expected position $\boldsymbol{u}_{ie}(k)$ of UAV $i$ in the formation is
\begin{equation}\label{13}
		e_{d,ie}(k)=\|\boldsymbol{u}_{ie}(k)-\boldsymbol{u}_i(k)\|,
\end{equation}
where $\|\cdot\|$ represents the Euclidean norm. The velocity error between UAV $i$ $\dot{\boldsymbol{u}}_i(k)=[\dot{x}_i,\dot{y}_i,\dot{z}_i]_k^\mathrm{T}$ and the virtual-leader $\dot{\boldsymbol{p}}_l(k)=[\dot{x}_l,\dot{y}_l,\dot{z}_l]_k^\mathrm{T}$ at time $k$ as follows.
\begin{equation}\label{14}
		e_{v,ie}(k)=\|\dot{\boldsymbol{p}}_l(k)-\dot{\boldsymbol{u}}_i(k)\|.
\end{equation}
The UAVs in the formation need to ensure the minimum formation safety distance $r_{\mathrm{min}}$.
\begin{equation}\label{15}
		r_{ij}(k)=\|\boldsymbol{u}_i(k)-\boldsymbol{u}_j(k)\|\geq r_{\mathrm{min}},
\end{equation}
where $i,j=1,2,\ldots,P$ and $i\neq j$. The minimum obstacle avoidance safety distance between UAV $i$ and obstacle $\boldsymbol{s}(k)=[x_s,y_s,z_s]_k^\mathrm{T}$ is $r_s$.
\begin{equation}\label{16}
		r_i(k)=\|\boldsymbol{u}_i(k)-\boldsymbol{s}(k)\|\geq r_s.
\end{equation}
The position, velocity and other relevant information of each UAV can be obtained from its own equipped global positioning system (GPS) and inertial navigation system (INS).

\subsection{Obstacle Dynamics}

Assume that during the UAV formation performing the virtual-leader path-following task, the dynamic obstacle will appear in 3D space, and the obstacle dynamics can be expressed by the following discrete-time dynamical model \cite{39papaioannou2020coordinated}.
\begin{equation}\label{17}
		\boldsymbol{o}(k)=\boldsymbol{\Phi o}(k-1)+\boldsymbol{\Gamma}\upsilon_k,
\end{equation}
where $\boldsymbol{o}=\big[\boldsymbol{s}^\mathrm{T},\dot{\boldsymbol{s}}^\mathrm{T}\big]^\mathrm{T}=\big[s_x,s_y,s_z,\dot{s}_x,\dot{s}_y,\dot{s}_z\big]^\mathrm{T}$ denotes the obstacle state, consisting of position and velocity vectors, and $\upsilon_k\sim\mathcal{N}(0,\boldsymbol{\Sigma}_v)$ denotes the perturbing acceleration, given by a zero mean multivariate Gaussian distribution with covariance matrix $\boldsymbol{\Sigma}_v$. The matrices $\boldsymbol{\Phi}$ and $\boldsymbol{\Gamma}$ are given by \cite{39papaioannou2020coordinated}
\begin{equation}\label{18}
	\begin{matrix}  
		\boldsymbol{\Phi}
    \end{matrix}
     =	
	\begin{bmatrix}  
		\mathbf{I}_3 & \triangle T\cdot\mathbf{I}_3 \\
		\mathbf{0}_3 & \mathbf{I}_3
	\end{bmatrix}
     ,
    \begin{matrix}  
     	\boldsymbol{\Gamma}
    \end{matrix}
     =	
	\begin{bmatrix}  
    	0.5\triangle T^2\cdot\mathbf{I}_3 \\
    	\triangle T\cdot\mathbf{I}_3
    \end{bmatrix}     ,
\end{equation}
where $\triangle T$ is the sampling period, $\mathbf{I}_3$ and $\mathbf{0}_3$ are the $3\times 3$ identity matrix and the zero matrix respectively. It is worth noting that the spatial obstacle state at the next time assumed depends only on the state at the previous time. In essence, its dynamics obeys the Markov property, as shown in \eqref{17}.

\section{Cramér-Rao Lower Bound for Obstacle Sensing Based on ISAC Signal}

To realize the spatial obstacle sensing, it is divided into two processes, obstacle range and radial velocity estimation based on ISAC signals and obstacle sensing based on information-level fusion. This section gives the principle of ISAC signal detection obstacle and its CRLB first, and obtains the range and radial velocity estimations between each UAV and obstacle. Then, the information detected by UAVs in formation is fused at the information-level to finally obtain the position and velocity estimates of the spatial obstacle, and the CRLB of UAV formation for obstacle sensing is derived based on the CRLB of ISAC signals from UAVs. It is worth noting that the obstacle sensing mentioned in this paper has the same meaning as the position and velocity estimation of the obstacle in 3D space.

\subsection{Obstacle Range and Radial Velocity Estimation Using ISAC Signal}

The UAVs in the formation use the 5G orthogonal frequency division multiplexing (OFDM) signals to realize information interaction and radar sensing. In \cite{40wei20225g}, the downlink positioning reference signal (PRS) of base station was employed for range and radial velocity estimation. In UAV networks, PRS may not be available because PRS exists in the downlink signal of base station, while the demodulation reference signal (DM-RS) is commonly available. Therefore, we follow the framework of \cite{40wei20225g} to provide a scheme for range and radial velocity estimation of the obstacle using DM-RS as the radar sensing signal and derive the CRLBs for range and velocity estimation.

We consider the non-uniform distribution of DM-RS in both the time domain and frequency domain. First, the DM-RS signal model is introduced, and the continuous time domain signal expression is given by \cite{chen2020performance}
\begin{equation}\label{25}
		x(t)=\sum\limits^{\lvert\mathcal{M}\rvert-1}\limits_{m=0}\sum\limits^{\lvert\mathcal{N}\rvert-1}\limits_{n=0}s_{m,n} e^{j2\pi f_nt} \mathrm{rect} \Big(\frac{t-mT_s}{T_s}\Big) ,
\end{equation}
where $\mathcal{M}$ is the set of OFDM symbols occupied by DM-RS in the time domain, and $\mathcal{N}$ is the set of subcarriers in the frequency domain. $s_{m,n}$ represents the modulated DM-RS symbol with OFDM symbol index $m$ and subcarrier index $n$ in the $\lvert\mathcal{M}\rvert$ OFDM symbols and $\lvert\mathcal{N}\rvert$ subcarriers carrying DM-RS. $T_s$ denotes the total duration of OFDM symbol, and $f_n$ denotes the frequency of the $n$-th subcarrier.

To use the radar detection function of DM-RS, it is necessary to divide the received modulation symbols $\boldsymbol{S}_{RX,i}[w,q]$ by the transmitted modulation symbols $\boldsymbol{S}_{TX,i}[w,q]$ in the receiver of the UAV $i$, namely \cite{40wei20225g}
\begin{equation}\label{26}
		(\boldsymbol{S}_i)_{w,q}=\displaystyle\frac{\boldsymbol{S}_{RX,i}[w,q]}{\boldsymbol{S}_{TX,i}[w,q]}=\xi e^{-j2\pi (q \triangle f \tau_i - w T_s f_{d,i})} ,
\end{equation}
where $w$ and $q$ are the actual indexes of the DM-RS in the total number of $M$ OFDM symbols and $N$ subcarriers, which implies that $w$ and $q$ are discontinuous and non-uniform. $(\boldsymbol{S}_i)_{w,q}$ is the channel information, $\xi$ is the attenuation factor, $\triangle f$ is the frequency spacing of subcarriers. $\tau_i=2r_i/c$ is the round-trip delay experienced by the signal transmission of UAV $i$, $r_i$ is the range from the UAV $i$ to the obstacle, and $c$ is the speed of light. $f_{d,i}=2\dot{r}_i f_c/c$ is the Doppler shift, $\dot{r}_i$ is the radial velocity, and $f_c$ is carrier frequency.

We adopt the radar signal processing algorithm in \cite{40wei20225g}. $N$-points inverse fast Fourier transform (IFFT) is performed in the frequency domain to obtain the peak $L_{\boldsymbol{s}_{i,w}}$ in the $w$-th column. Similarly, $M$-points fast Fourier transform (FFT) is performed in the time domain to obtain the $q$-th row peak index $L_{\boldsymbol{s}_{i,q}}$.
\begin{equation}\label{27}
		r(l_i)=\mathsf{IFFT}(\boldsymbol{s}_{i,w})=\sum\limits^{\lvert\mathcal{N}\rvert-1}\limits_{n=0} e^{-j2\pi \boldsymbol{q}_n \triangle f \tau_i} e^{j2\pi l_i \frac{n}{\lvert\mathcal{N}\rvert}},
\end{equation}
\begin{equation}\label{28}
		v(d_i)=\mathsf{FFT}(\boldsymbol{s}_{i,q})=\sum\limits^{\lvert\mathcal{M}\rvert-1}\limits_{m=0} e^{j2\pi \boldsymbol{w}_m T_s f_{d,i}} e^{-j2\pi d_i \frac{m}{\lvert\mathcal{M}\rvert}},
\end{equation}
where $\boldsymbol{w}_m$ represents the actual index of the $m$-th DM-RS symbol extracted from the set formed by $w$, and $\boldsymbol{q}_n$ represents the actual index of the $n$-th subcarrier extracted from the set formed by $q$. $l_i\in \{0,1,\ldots,\lvert\mathcal{N}\rvert-1\}$ and $d_i\in \{0,1,\ldots,\lvert\mathcal{M}\rvert-1\}$ are the result indexes of IFFT and FFT. $\boldsymbol{s}_{i,q}$ and $\boldsymbol{s}_{i,w}$ denote the extraction of a row or a column from $(\boldsymbol{S}_i)_{w,q}$.

The peaks occur respectively when the sum of the squares of the phases of all $n$ or $m$ in \eqref{27} and \eqref{28} are minimized. After obtaining the time-domain and frequency-domain peak indexes, the range and radial velocity estimates between UAV $i$ and the obstacle can be calculated by
\begin{equation}\label{29}
		r_i= \frac{L_{\boldsymbol{s}_{i,w}} c \sum\nolimits^{\lvert\mathcal{N}\rvert-1}\nolimits_{n=0} n^2}{2 \lvert\mathcal{N}\rvert \triangle f \sum\nolimits^{\lvert\mathcal{N}\rvert-1}\nolimits_{n=0} n \boldsymbol{q}_n},
\end{equation}
\begin{equation}\label{30}
		\dot{r}_i= \frac{L_{\boldsymbol{s}_{i,q}} c \sum\nolimits^{\lvert\mathcal{M}\rvert-1}\nolimits_{m=0} m^2}{2 \lvert\mathcal{M}\rvert T_s f_c \sum\nolimits^{\lvert\mathcal{M}\rvert-1}\nolimits_{m=0} m \boldsymbol{w}_m}.
\end{equation}

\subsection{CRLB for Obstacle Range and Radial Velocity Estimation}

UAVs in the formation send OFDM signals and receive echo signals from the obstacle for range and radial velocity estimation. Since the modulation symbol is known to the radar receiver and the noise obeys a one-dimensional Gaussian distribution, the DM-RS radar detection signal in the UAV $i$ after phase-by-phase rotation is \cite{40wei20225g}
\begin{equation}\label{31}
		z_{mn,i}=\xi A_{mn,i} e^{-j2\pi \boldsymbol{q}_n \triangle f \tau_i} e^{j2\pi \boldsymbol{w}_m T_s f_{d,i}} + w_{mn,i} ,
\end{equation}
where $A_{mn,i}$ denotes the transmitted symbol amplitude of UAV $i$, $w_{mn,i}$ is an additive white Gaussian noise (AWGN) with zero mean and $\sigma^2$ variance. The unknown parameter $\theta_i=(\tau_i,f_{d,i})$ include the round-trip delay $\tau_i$ and the Doppler shift $f_{d,i}$. 

We refer to the method in \cite{40wei20225g} for deriving the CRLB for uniform PRS based radar sensing and propose our CRLB based on non-uniform DM-RS for range and radial velocity estimation. The minimized log likelihood function is
\begin{equation}\label{32}
	\begin{small}
	\begin{aligned}  
		 L(z_i|\tau_i,f_{d,i}) &= \ln \left[(2\pi\sigma^2)^{-\frac{MN}{2}} e^{-\frac{1}{2\sigma^2}\sum\limits_m\sum\limits_n \left|z_{mn,i}-s_{mn,i} \right|^2}\right] \\
		 & = -\frac{MN\ln(2\pi\sigma^2)}{2}-\frac{\sum\limits_m\sum\limits_n \left|z_{mn,i}-g_{mn,i} \right|^2}{2\sigma^2} ,
	\end{aligned}
    \end{small}
\end{equation}
where
\begin{equation}\label{33}
		g_{mn,i}=\xi A_{mn,i} e^{-j2\pi \boldsymbol{q}_n \triangle f \tau_i} e^{j2\pi \boldsymbol{w}_m T_s f_{d,i}}.
\end{equation}

Further, the second-order Fisher information matrix (FIM) $\boldsymbol{J}_i$ is
\begin{equation}\label{34}
		\boldsymbol{J}_i
    =-
   	\begin{bmatrix}  
   	    \mathbb{E}\left(\frac{\partial^2 L(z_i|\tau_i,f_{d,i})}{\partial\tau_i^2}\right) 
   	    &  \mathbb{E}\left(\frac{\partial^2 L(z_i|\tau_i,f_{d,i})}{\partial\tau_i \partial f_{d,i}}\right)\\
   	    \mathbb{E}\left(\frac{\partial^2 L(z_i|\tau_i,f_{d,i})}{\partial f_{d,i} \partial\tau_i }\right)
   	    &  \mathbb{E}\left(\frac{\partial^2 L(z_i|\tau_i,f_{d,i})}{\partial f_{d,i}^2 }\right)
    \end{bmatrix}.
\end{equation}

The CRLB matrix of the time delay and Doppler shift estimates of UAV $i$ for obstacle is the inverse of the Fisher matrix $\boldsymbol{J}_i$, that is
\begin{equation}\label{35}
	\begin{bmatrix}  
		CRLB_{\tau_i} &  
		CRLB_{\tau_i,f_{d,i}} \\
		CRLB_{f_{d,i},\tau_i} & 
		CRLB_{f_{d,i}}
	\end{bmatrix}
    =
	\begin{matrix}  
        \boldsymbol{J}_i^{-1}
    \end{matrix}.
\end{equation}

Then the CRLB for round-trip delay and Doppler shift estimation based on UAV $i$ can be obtained.
\begin{equation}\label{36}
	\begin{matrix}  
		CRLB_{\tau_i} = \frac{\mathbb{E}(\frac{\partial^2 L}{\partial f_{d,i}^2})}{
		\mathbb{E}(\frac{\partial^2 L}{\partial \tau_i^2})\mathbb{E}(\frac{\partial^2 L}{\partial f_{d,i}^2})
		- \mathbb{E}(\frac{\partial^2 L}{\partial \tau_i f_{d,i}}) 
     	\mathbb{E}(\frac{\partial^2 L}{\partial f_{d,i} \tau_i }) },
	\end{matrix}
\end{equation}
\begin{equation}\label{37}
	\begin{matrix}  
		CRLB_{f_{d,i}} = \frac{\mathbb{E}(\frac{\partial^2 L}{\partial \tau_i^2})}{
			\mathbb{E}(\frac{\partial^2 L}{\partial \tau_i^2})\mathbb{E}(\frac{\partial^2 L}{\partial f_{d,i}^2})
			- \mathbb{E}(\frac{\partial^2 L}{\partial \tau_i f_{d,i}}) 
			\mathbb{E}(\frac{\partial^2 L}{\partial f_{d,i} \tau_i }) } .
	\end{matrix}
\end{equation}

According to the parameters transformation relationship between range and time delay, radial velocity and Doppler shift, the CRLB for range and radial velocity estimation between UAV $i$ and obstacle can be obtained as follows, where $SNR$ represents signal-to-noise ratio.
\begin{equation}\label{38}
	\begin{aligned}  
		&CRLB_{r_i} = \frac{c^2}{4} CRLB_{\tau_i} \\
		&= \frac{c^2 \sum\limits_{m,n} \boldsymbol{w}^2_m} {16 \xi^2 \pi^2 SNR \triangle f^2 \sum\limits_{m,n} \boldsymbol{q}^2_n \sum\limits_{m,n} \boldsymbol{w}^2_m - \sum\limits_{m,n} \boldsymbol{q}^2_n \boldsymbol{w}^2_m},
	\end{aligned}
\end{equation}
\begin{equation}\label{39}
	\begin{aligned}  
		&CRLB_{v_i} = \frac{c^2}{4f_c^2} CRLB_{f_{d,i}} \\
		&= \frac{c^2 \sum\limits_{m,n} \boldsymbol{q}^2_n} {16 \xi^2 \pi^2 SNR f_c^2 T_s^2 \sum\limits_{m,n} \boldsymbol{q}^2_n \sum\limits_{m,n} \boldsymbol{w}^2_m - \sum\limits_{m,n} \boldsymbol{q}^2_n \boldsymbol{w}^2_m}.
	\end{aligned}
\end{equation}
	
\subsection{Obstacle Sensing Strategy}

The UAV formation uses cooperative sensing to estimate the position and velocity of the spatial obstacle, assuming that UAVs in the formation have strict clock synchronization. Each UAV sends ISAC signals and receives echoes independently, and uses \eqref{29} and \eqref{30} to estimate the range $r_i$ and the radial velocity $\dot{r}_i$.

Integrate the range and radial velocity estimates of each UAV to the obstacle into the MUAV for information-level fusion, and obtain the joint position and velocity estimates of the obstacle. Using both range and radial velocity for the obstacle position and velocity estimation improves the accuracy and requires fewer UAVs than the methods that use only distance or radial velocity \cite{42yeredor2010joint}.

To jointly estimate the position and velocity of the spatial obstacle, first, the range difference $r_{i1}$ and the radial velocity difference $\dot{r}_{i1}$ between the obstacle to UAV $i$ and to MUAV are calculated as \cite{43ho2004accurate}
\begin{equation}\label{40}
		r_{i1} = r_i - r_1 + n_{i1} = \| \boldsymbol{s} - 
		\boldsymbol{u}_i \| - \| \boldsymbol{s} - 
		\boldsymbol{u}_1 \| + n_{i1},
\end{equation}
\begin{equation}\label{41}
	\begin{matrix}  
		\dot{r}_{i1} = \dot{r}_i - \dot{r}_1 + \dot{n}_{i1} = \frac{(\boldsymbol{s} - 
			\boldsymbol{u}_i)^\mathrm{T} (\dot{\boldsymbol{s}} - \dot{\boldsymbol{u}}_i)}{r_i} - \frac{(\boldsymbol{s} - 
			\boldsymbol{u}_1)^\mathrm{T} (\dot{\boldsymbol{s}} - \dot{\boldsymbol{u}}_1)}{r_1} + \dot{n}_{i1} ,
	\end{matrix} 
\end{equation}
where $n_{i1}$ and $\dot{n}_{i1}$ are error terms, and the UAV formation cooperative sensing error vector $\boldsymbol{n}=[n_{21},\ldots,n_{P1},$ $\dot{n}_{21},\ldots,\dot{n}_{P1}]^\mathrm{T}$ obeys a Gaussian distribution with zero mean and $\boldsymbol{Q}$ covariance matrix. The error vector ignoring the second-order error term as follows \cite{43ho2004accurate}.
\begin{equation}\label{42}
	\begin{aligned}  
		\boldsymbol{\varepsilon}_r = r_{i1}^2 + 2r_{i1}r_1 - \boldsymbol{u}_i^\mathrm{T}\boldsymbol{u}_i + \boldsymbol{u}_1^\mathrm{T}\boldsymbol{u}_1 + 2(\boldsymbol{u}_i-\boldsymbol{u}_1)^\mathrm{T}\boldsymbol{s},
	\end{aligned}
\end{equation}
\begin{equation}\label{43}
	\begin{aligned}  
		\boldsymbol{\varepsilon}_v = & 2(\dot{\boldsymbol{u}}_i-\dot{\boldsymbol{u}}_1)^\mathrm{T}\boldsymbol{s} + 2(\boldsymbol{u}_i-\boldsymbol{u}_1)^\mathrm{T}\dot{\boldsymbol{s}}  \\  & + 2\dot{r}_{i1}r_1 + 2r_{i1}\dot{r}_1 - 2(\dot{\boldsymbol{u}}_i^\mathrm{T}\boldsymbol{u}_i - \dot{\boldsymbol{u}}_1^\mathrm{T}\boldsymbol{u}_1 - r_{i1}\dot{r}_{i1}).
	\end{aligned}
\end{equation}
where $\boldsymbol{\varepsilon}_r$ and $\boldsymbol{\varepsilon}_v$ are the range and radial velocity errors between UAV $i$ and the obstacle, respectively.

The objective is to calculate the obstacle position $\boldsymbol{s}$ and velocity $\dot{\boldsymbol{s}}$ sensed by the UAV formation. Considering the variable to be solved is $\boldsymbol{\zeta}=[\boldsymbol{s}^\mathrm{T},r_1,\dot{\boldsymbol{s}}^\mathrm{T},\dot{r_1}]^\mathrm{T}$. Construct the error vector $\boldsymbol{\varepsilon}_1=[\boldsymbol{\varepsilon}_r,\boldsymbol{\varepsilon}_v]^\mathrm{T}$ according to \eqref{42}, \eqref{43}, and use the two-step weighted least squares (TWLS) method in \cite{43ho2004accurate} to calculate $\boldsymbol{s}$ and $\dot{\boldsymbol{s}}$ as
\begin{equation}\label{44}
	\begin{matrix}  
		\boldsymbol{s} = \boldsymbol{U} \Big [\sqrt{\boldsymbol{\zeta}_2(1)}, \sqrt{\boldsymbol{\zeta}_2(2)}, \sqrt{\boldsymbol{\zeta}_2(3)}\Big]^\mathrm{T} +\boldsymbol{u}_1,
	\end{matrix}
\end{equation}
\begin{equation}\label{45}
	\begin{matrix}  
		\dot{\boldsymbol{s}} = \boldsymbol{U} \Big [\frac{\boldsymbol{\zeta}_2(4)}{\sqrt{\boldsymbol{\zeta}_2(1)}},  \frac{\boldsymbol{\zeta}_2(5)}{\sqrt{\boldsymbol{\zeta}_2(2)}},  \frac{\boldsymbol{\zeta}_2(6)}{\sqrt{\boldsymbol{\zeta}_2(3)}} \Big]^\mathrm{T} +\dot{\boldsymbol{u}}_1,
	\end{matrix}
\end{equation}
\begin{equation}\label{46}
	\begin{matrix}  
		\boldsymbol{U} = \mathrm{diag}\{\mathrm{sgn}(\boldsymbol{\zeta}_2(1:3)-\boldsymbol{u}_1)\},
	\end{matrix}
\end{equation}
\begin{equation}\label{47}
	\begin{matrix}  
		\boldsymbol{\zeta}_2
	\end{matrix}
     = 
	\begin{bmatrix}  
		(\boldsymbol{s}-\boldsymbol{u}_1) \odot (\boldsymbol{s}-\boldsymbol{u}_1) \\
		(\dot{\boldsymbol{s}}-\dot{\boldsymbol{u}}_1) \odot (\boldsymbol{s}-\boldsymbol{u}_1)
	\end{bmatrix}.
\end{equation}
where $\mathrm{diag}\{\cdot\}$ denotes the diagonal matrix, $\mathrm{sgn}(\cdot)$ denotes the sign function, and $\odot$ denotes the Hadamard product.

Meanwhile, UAVs employing ISAC signals significantly reduce communication overhead compared to traditional systems that separate communication and sensing, which can be attributed to the integrated features of ISAC signals.

Firstly, UAVs utilize ISAC signals, which possess the fusion capability of both communication and sensing. This characteristic avoids the sequential transmission of radar and communication signals, resulting in reduced communication overhead. In traditional systems, each UAV must individually transmit radar signals and receive echoes for obstacle detection and estimation. Subsequently, AUAVs transmit measurement data to the MUAV for fusion, obtaining estimates of the obstacle position and velocity. Finally, the MUAV sends VFTs back to each AUAV for coordinate enhanced sensing or obstacle avoidance. This process incurs significant communication overhead when communication and sensing are separate. In contrast, the OFDM signals transmitted by AUAVs in the ISAC system encompass both radar signals for obstacle range and radial velocity estimation, as well as the estimated obstacle information conveyed to the MUAV. Similarly, the OFDM signals from the MUAV contain the VFTs for the AUAVs in the next time step, thus eliminating the need for additional communication overhead.

Specifically, in a UAV formation employing ISAC technology, communication overhead is reduced by a factor of $2C_t(P-1)$ compared to a traditional system with separate communication and sensing. Where $P-1$ represents the number of AUAVs, and $C_t$ is the communication overhead of an individual UAV in the traditional system. This efficiency improvement is crucial for enhancing UAV communication performance and optimizing resource management, especially in scenarios where multiple UAVs collaborate.

\subsection{CRLB for Obstacle Sensing}

The CRLB for obstacle sensing is related to the position and velocity of UAVs and the obstacle, which gives the lower bound of the error covariance matrix in the unbiased estimator \cite{42yeredor2010joint}. In \cite{45del2012achievable}, the CRLB for achievable positioning was derived using the CRLB for range estimation of OFDM signals. With the difference that in this subsection, we derive the CRLB for obstacle sensing in UAV formation based on the CRLBs for range and radial velocity estimation of an individual UAV in \eqref{38} and \eqref{39}, along with the fused sensing strategy.

Assume that the parameter to be estimated is $\boldsymbol{\gamma}=\big[\boldsymbol{u}^\mathrm{T},\dot{\boldsymbol{u}}^\mathrm{T},\boldsymbol{s}^\mathrm{T},\dot{\boldsymbol{s}}^\mathrm{T}\big]^\mathrm{T}$, which includes the position and velocity vectors of UAVs in the formation and obstacle, and for the unbiased estimate $\hat{\boldsymbol{\gamma}}$, we have
\begin{equation}\label{48}
	\begin{aligned}  
		 \mathbb{E} \big[(\hat{\boldsymbol{\gamma}}-\boldsymbol{\gamma}) (\hat{\boldsymbol{\gamma}}-\boldsymbol{\gamma})^\mathrm{T}\big] \geq \boldsymbol{J}^{-1} = \boldsymbol{CRLB}_{PV},
	\end{aligned}
\end{equation}
where $\boldsymbol{CRLB}_{PV}$ is the CRLB for obstacle sensing, the estimated values of obstacle state and the UAVs state vectors obey the Gaussian distributions and are independent of each other. The probability density function of the data vector $\boldsymbol{\mu}=\big[\boldsymbol{r}_{rv}^\mathrm{T},\boldsymbol{u}_{rv}^\mathrm{T}\big]^\mathrm{T}$ is
\begin{equation}\label{49}
	\begin{aligned}  
		\ln f(\boldsymbol{\mu}; \boldsymbol{\gamma}) = &- \frac{1}{2} \ln ((2\pi)^{2P-2} |\boldsymbol{Q}|) \\ &- \frac{1}{2} (\boldsymbol{r}_{rv} - \boldsymbol{r}_{rv}^o)^\mathrm{T} \boldsymbol{Q}^{-1}  (\boldsymbol{r}_{rv} - \boldsymbol{r}_{rv}^o),
	\end{aligned}
\end{equation}
where $\ln f(\boldsymbol{\mu}; \boldsymbol{\gamma})$ is Gaussian with mean $\boldsymbol{r}_{rv}^o$ and covariance matrix $\boldsymbol{Q}$, $\boldsymbol{r}_{rv}=[r_{21},\ldots,r_{P1},\dot{r}_{21},\ldots,\dot{r}_{P1}]^\mathrm{T}$ is the vector consisting of the position difference and radial velocity difference from obstacle to AUAVs and obstacle to MUAV, which is obtained by each AUAV integrating its own range and radial velocity estimation to MUAV. $\boldsymbol{r}_{rv}^o=[r_{21}^o,\ldots,r_{P1}^o,\dot{r}_{21}^o,\ldots,\dot{r}_{P1}^o]^\mathrm{T}$ is the true value. $\boldsymbol{u}_{rv}=\big[\boldsymbol{u}^\mathrm{T},\dot{\boldsymbol{u}}^\mathrm{T}\big]^\mathrm{T}$ is the position and velocity vector of the UAVs in the formation. Then the FIM of $\boldsymbol{\gamma}$ is \cite{43ho2004accurate}
\begin{equation}\label{50}
	\begin{aligned}  
		\boldsymbol{J}(\boldsymbol{\gamma}) = -\mathbb{E} \bigg[\frac{\partial^2 \ln f(\boldsymbol{\mu}; \boldsymbol{\gamma})}{\partial\boldsymbol{\gamma}\partial\boldsymbol{\gamma}^\mathrm{T}}\bigg] = \bigg(\frac{\partial\boldsymbol{r}_{rv}^o}{\partial\boldsymbol{\gamma}}\bigg)^\mathrm{T} \boldsymbol{Q}^{-1}\bigg(\frac{\partial\boldsymbol{r}_{rv}^o}{\partial\boldsymbol{\gamma}} \bigg),
	\end{aligned}
\end{equation}
where 
\begin{equation}\label{51}
	\begin{matrix}  
		\displaystyle\frac{\partial\boldsymbol{r}_{rv}^o}{\partial\boldsymbol{\gamma}} =
	\end{matrix}
	\begin{bmatrix}  
		\boldsymbol{A} & \boldsymbol{0}_{(P-1)\times 3} \\
		\boldsymbol{B} & \boldsymbol{A}
	\end{bmatrix},
\end{equation}
\begin{equation}\label{52}
	\begin{matrix}  
		\boldsymbol{A} =
	\end{matrix}
	\begin{bmatrix}  
		\frac{(\boldsymbol{s} - 
			\boldsymbol{u}_2)^\mathrm{T}}{r_2^o} - \frac{(\boldsymbol{s} - 
			\boldsymbol{u}_1)^\mathrm{T}}{r_1^o} \\
		\vdots \\
		\frac{(\boldsymbol{s} - 
			\boldsymbol{u}_P)^\mathrm{T}}{r_P^o} -  \frac{(\boldsymbol{s} - 
			\boldsymbol{u}_1)^\mathrm{T}}{r_1^o} 
	\end{bmatrix},
\end{equation}
\begin{equation}\label{53}
	\begin{matrix}  
		\boldsymbol{B} =
	\end{matrix}
	\begin{bmatrix}  
		\Big(\frac{(\dot{\boldsymbol{s}} - 
			\dot{\boldsymbol{u}}_2)^\mathrm{T}}{r_2^o}-
		\frac{\dot{r}_2^o(\boldsymbol{s} - 
			\boldsymbol{u}_2)^\mathrm{T}}{r_2^{o2}}\Big) - \Big(\frac{(\dot{\boldsymbol{s}} - \dot{\boldsymbol{u}}_1)^\mathrm{T}}{r_1^o}-
		\frac{\dot{r}_1^o(\boldsymbol{s} - 
			\boldsymbol{u}_1)^\mathrm{T}}{r_1^{o2}}\Big) \\
		\vdots \\
		\Big(\frac{(\dot{\boldsymbol{s}} - 
			\dot{\boldsymbol{u}}_P)^\mathrm{T}}{r_P^o}-
		\frac{\dot{r}_P^o(\boldsymbol{s} - 
			\boldsymbol{u}_P)^\mathrm{T}}{r_P^{o2}}\Big) - \Big(\frac{(\dot{\boldsymbol{s}} - \dot{\boldsymbol{u}}_1)^\mathrm{T}}{r_1^o}-
		\frac{\dot{r}_1^o(\boldsymbol{s} - 
			\boldsymbol{u}_1)^\mathrm{T}}{r_1^{o2}}\Big)
	\end{bmatrix}.
\end{equation}
In \eqref{52} and \eqref{53}, $r_i^o$ and $\dot{r}_i^o$ are the true value of the range and radial velocity between the UAV $i$ and the obstacle, which cannot be obtained in practice, and the arithmetic mean of multiple measurements is used instead. The covariance matrix $\boldsymbol{Q}$ is
\begin{equation}\label{54}
	\begin{aligned}  
		\boldsymbol{Q} = diag\{\boldsymbol{Q}_r,\boldsymbol{Q}_v \},
	\end{aligned}
\end{equation}
\begin{equation}\label{55}
	\begin{matrix}  
		\boldsymbol{Q}_\Lambda =
	\end{matrix}
	\begin{bmatrix}  
	\sigma_{\Lambda,1}^2+\sigma_{\Lambda,2}^2 & \sigma_{\Lambda,1}^2 & \cdots & \sigma_{\Lambda,1}^2 \\
	\sigma_{\Lambda,1}^2 & \sigma_{\Lambda,1}^2+\sigma_{\Lambda,3}^2 & \cdots & \sigma_{\Lambda,1}^2
	\\
	\vdots & \vdots & \ddots & \vdots
	\\
	\sigma_{\Lambda,1}^2 & \sigma_{\Lambda,1}^2 & \cdots & \sigma_{\Lambda,1}^2 + \sigma_{\Lambda,P}^2
    \end{bmatrix},
\end{equation}
where $\Lambda\in\{r,v\}$. Being $\sigma_{r,i}^2$ and $\sigma_{v,i}^2$ the variance, which are defined by  $CRLB_{r_i}$ and $CRLB_{v_i}$, respectively 
\cite{45del2012achievable}, as shown in \eqref{38} and \eqref{39}.

Consider $\boldsymbol{CRLB}_{PV}^{i:j}$ as a square matrix of size $(j-i+1)\times(j-i+1)$, consisting of the row $i$ and column $i$ to row $j$ and column $j$ of $\boldsymbol{CRLB}_{PV}$. The minimum position and velocity errors of the obstacle are finally calculated as
\begin{equation}\label{56}
	\begin{aligned}  
		\varepsilon_P = \sqrt{\mathrm{tr}(\boldsymbol{CRLB}_{PV}^{1:3})} = \sqrt{\mathrm{tr}(\boldsymbol{CRLB}_{P})},
	\end{aligned}
\end{equation}
\begin{equation}\label{57}
	\begin{aligned}  
		\varepsilon_V = \sqrt{\mathrm{tr}(\boldsymbol{CRLB}_{PV}^{4:6})} = \sqrt{\mathrm{tr}(\boldsymbol{CRLB}_{V})},
	\end{aligned}
\end{equation}
where $\mathrm{tr}(\cdot)$ is the trace of the matrix.

\section{Proposed Path-Following, Obstacle Sensing, and Obstacle Avoidance Algorithms}

In this section, we solve the three subtasks in the UAV formation cooperative perform path-following task in 3D space, that is, path-following subtask, obstacle sensing subtask, and obstacle avoidance subtask. We propose the DRL-based UAV formation path-following algorithm, the variable formation enhanced obstacle position estimation algorithm, and the online obstacle avoidance algorithm based on DRL path-following. In addition, we give the corresponding complexity analysis of the algorithms.

\subsection{UAV Formation Path-Following Based on DRL}

DRL path-following training is a process of continuous iterative interaction with the environment. UAVs obtain observable states from the environment and make reasonable action decisions. The reward of the action is used to train and update the model. The model gradually stabilizes during the continuous iterative repetition until it can take the desired actions to complete the path-following subtasks.

In the fixed weight DRL path-following (FWPF) algorithm based on formation training represented by \cite{zhao2021usv}, two aspects are worthy of attention. On the one hand, the velocity error between UAVs and the target is not considered in the reward function using only the distance error with a fixed weight. To obtain higher path-following accuracy, we develop a reward function with adaptive weights that takes into account both distance and velocity errors. On the other hand, the core of the path-following algorithm based on UAV formation training lies in the control of the formation. By adding the effect of formation error to the reward function, the consistency of UAVs is realized, and each UAV selects its position in the formation according to the virtual-leader position and formation restrictions. In this paper, we design a path-following algorithm trained exclusively for an individual UAV. This algorithm demonstrates low model complexity, rapid convergence, and strong scalability, enabling easy adjustment of the number of UAVs. Specifically, the AUAVs do not need to know the virtual-leader position and the formation shape, but only follow the VFT position received from the MUAV.

The state space of UAV $i$ in the formation is denoted as $s_i=(x_i,y_i,z_i,\chi_i,\gamma_i,\dot{x}_i,\dot{y}_i,\dot{z}_i,x_{ie},y_{ie},z_{ie},e_{d,ie},e_{v,ie})$, where $x_i,y_i,z_i$ and $\dot{x}_i,\dot{y}_i,\dot{z}_i$ are position and velocity of UAV $i$ in 3D space. $\chi_i,\gamma_i$ are the track angle and heading angle, $x_{ie},y_{ie},z_{ie}$ are the VFT assigned to UAV $i$ by MUAV, and $e_{d,ie},e_{v,ie}$ are distance and velocity errors. Using the DRL algorithm, the corresponding decision-making actions $a_i=(v_{xi},v_{yi},v_{zi})$, i.e., control velocity of UAV $i$, can be output under the input of UAV $i$ observed states. It is worth noting that action decisions are distributed in a continuous space.

Distinguished from the traditional DRL path-following strategy, we jointly consider the effects of the distance error of UAVs and VFTs and the velocity error of UAVs and virtual-leader in the reward function, and the weights can be adaptively adjusted to improve the path-following accuracy. To avoid the sparse reward, a continuous reward function used in this paper, is shown as follows.
\begin{equation}\label{19}
	{\mathscr{R}} = -\omega_1 e_{d,ie} -\omega_2 e_{v,ie},\ \omega_2 \propto (1/e_{d,ie}),
\end{equation}
where $\omega_1,\omega_2\in[0,1]$, and $\sum\nolimits_{o=1}^{2}\omega_o=1$. The $\omega_1$ and $\omega_2$ are the adaptive weights of distance and velocity factors respectively. $e_{d,ie}$ and $e_{v,ie}$ are calculated using \eqref{13}, \eqref{14}. The symbol $\propto$ signifies a direct proportionality.

In the chasing phase, as shown in Fig. \ref{fig_1} $t_0\sim t_1$, UAVs are distant from the virtual-leader, and $e_{d,ie}$ dominates in the reward function to make the UAVs move toward the VFTs at a large speed, so as to promote rapid formation. The adaptive $\omega_2$ increases gradually as $e_{d,ie}$ decreases until the maximum value. In the path-following phase (Fig. \ref{fig_1} after $t_1$), $e_{d,ie}$ and $e_{v,ie}$ are mutually constrained to ensure that the following distance error and the velocity error at the current moment are simultaneously minimized, contributing to the precise following at the next time. Since our obstacle avoidance scheme is performed online on the trained DRL model, $e_{v,ie}$ cannot be used as the dominant of the following phase to improve accuracy. This is because although the DRL path-following subtask is projected into the null space of the obstacle avoidance subtask (Section V) while performing the obstacle avoidance, the path-following subtask still completes partially. If $e_{v,ie}$ dominates the reward function in the path-following phase, UAVs blindly follow the velocity of the virtual-leader instead of accelerating to catch up with it after performing obstacle avoidance, adversely affecting the path-following performance. Note that the dominance of $e_{v,ie}$ in the reward function implies it contributes significantly to the reward value, rather than having a high weight $\omega_2$.

The path-following model in this paper adopts a DRL algorithm based on DDPG, and the proposed path-following model for VFTs aims to learn a shared policy $a=\mu(s|\varpi^{\mu})$ that can be used for all UAVs in the formation. The actor network output strategy is updated using sampled gradient \cite{zhao2021usv}
\begin{equation}\label{20}
	\nabla_{\theta^\mu}\mu | _{s_i} \approx \frac{1}{B} \sum_i \nabla_a Q(s,a|\theta^Q) | _{s=s_i,a=\mu(s_i)} \nabla_{\theta^\mu} \mu (s|\theta^\mu) |_{s_i}.
\end{equation}

In the network training process, randomly sampled small batches of data are used for training. The mini-batch size is $B$, $\mu (s|\theta^\mu)$ is the action in the state $s$, $Q(s,a|\theta^Q)$ is the evaluation value of the state-action pair $(s,a)$. The gradient descent algorithm maximizes the cumulative expected return and updates the actor network parameters.

For the critic network, the network parameter $\theta^Q$ is updated by minimizing the difference between the evaluated and expected values through a gradient descent algorithm \cite{zhao2021usv},
\begin{equation}\label{21}
	L(\theta^Q)=\frac{1}{B}\sum_i \big(y_i - Q(s_i,a_i|\theta^Q)\big)^2,
\end{equation}
\begin{equation}\label{22}
	y_i = r_i + \gamma Q^{'} \big(s_{i+1},\mu^{'}(s_{i+1}|\theta^{\mu^{'}})| \theta^{Q^{'}}\big),
\end{equation}
where $\gamma$ is the discount factor, $\mu^{'}(\cdot|\theta^{\mu^{'}})$ and $Q^{'}(\cdot|\theta^{Q^{'}})$ are the target actor network and target critic network respectively, and $y_i$ is the target value of state-action pair $(s_{i+1},a_{i+1})$.

The proposed adaptive weights DRL path-following (AWPF) algorithm is shown in Algorithm 1.

\renewcommand{\algorithmicrequire}{\textbf{Input:}}  
\renewcommand{\algorithmicensure}{\textbf{Output:}} 
\begin{algorithm}[!t]
	\caption{AWPF Algorithm Based on DDPG} 
	\label{alg::conjugateGradient}
	\begin{algorithmic}[1]
		\Require
		$x_i,y_i,z_i,\chi_i,\gamma_i,\dot{x}_i,\dot{y}_i,\dot{z}_i,x_{ie},y_{ie},z_{ie},e_{d,ie},e_{v,ie}$
		\Ensure
		$v_{xi},v_{yi},v_{zi}$
		\State Initialize the scenario parameters;
		\State Randomly initialize critic network $Q(s,a|\theta^Q)$ and actor $\mu(s|\theta^\mu)$ with weights $\theta^Q$ and $\theta^\mu$;
		\State Initialize target network $Q^{'}$ and $\mu^{'}$ with $\theta^{Q^{'}} $ and $\theta^{\mu^{'}}$;
		\State Initialize replay buffer $R$;
		\For {$i=1,2,\ldots$}
		\State Initialize a random process $\mathcal{N}$ for action exploration;
		\State Reset initial observation state $s_1$;
		\For {$t=1,2,\ldots$,T}
		\State Select action $a_t=\mu(s_t|\theta^\mu)+\mathcal{N}_t$ according to the \Statex \qquad  \quad current policy and exploration noise;
		\State Execute action $a_t$, obtain reward $r_t$ and observe 
		\Statex \qquad  \quad new state $s_{t+1}$;
		\State Store transition $(s_t,a_t,r_t,s_{t+1})$ in $R$ and 
		\Statex \qquad  \quad randomly sample a minibatch of $B$ transitions 
		\Statex \qquad  \quad $(s_i,a_i,r_i,s_{i+1})$.
		\State Update network parameters $\theta^Q$ and $\theta^\mu$ using 
		\Statex \qquad  \quad gradient descent method in \eqref{20}, \eqref{21};
		\State Update the target network using $\theta^Q$ and $\theta^\mu$.
		\EndFor
		\EndFor
	\end{algorithmic}
\end{algorithm}

\subsection{Complexity Analysis of the AWPF Algorithm}

In this section, we present the complexity analysis of the AWPF algorithm. Firstly, we evaluate the complexity of the AWPF decision time based on basic operations. We assume that addition or subtraction between two $\kappa$-bit numbers requires $\kappa$-bit operations, while multiplication or division requires $\kappa^2$-bit operations \cite{huxin2023}, and the UAV formation path-following parameters are represented with $\kappa$-bit precision.

In the AWPF algorithm, the complexity of each iteration is firstly influenced by MUAV computing VFTs for $P-1$ AUAVs, as calculated according to \eqref{10}, \eqref{11}, and \eqref{12}, resulting in the complexity of $\Psi_{\texttt{VFT}}=4(P-1)(\kappa^2+\kappa+1)$. Subsequently, the computation of distance and velocity errors in \eqref{13} and \eqref{14} contributes complexity $\Psi_{\texttt{e}}=P(3\kappa^2+5\kappa+1)$ separately. The computational complexity of the AWPF algorithm primarily depends on the complexity of the neural networks in DRL, where the number of multiply-accumulate operations (MACCs) in the fully connected layers is $S_{l-1}\times S_{l}$. The complexity of the actor network with $L_{n}$ layers is $\Psi_{\texttt{DRL}}=P\sum_{l=1}^{L_{n}}S_{l-1}\times S_{l} \times \kappa^2$, where $S_{l-1}$ and $S_{l}$ are the input and output dimension of the $l$-th network respectively. The computational load of activation functions is significantly lower than that of matrix operations in fully connected layers and is typically negligible. Therefore, the complexity of the AWPF algorithm can be expressed as
\begin{equation}\label{23} 
	\Psi_{\texttt{A}} = \Psi_{\texttt{VFT}}+2\Psi_{\texttt{e}}+\Psi_{\texttt{DRL}}
	\approx P (\sum_{l=1}^{L_{n}}S_{l-1}\times S_{l})\kappa^2 +14 P \kappa.
\end{equation}
It reveals that both the number of UAVs in the formation and the depth of neural networks in the DRL algorithm, as well as the input and output dimensions of the fully connected layers, have a linear impact on the complexity of the AWPF algorithm.

\subsection{Variable Formation Enhanced Obstacle Position Estimates}

At time $k$, the UAV formation observes the obstacle and obtains the estimated position and velocity, from which the estimated obstacle position at time $k+1$ can be calculated. Assuming a positioning accuracy threshold of $\zeta_{CRLB}^P$, we can calculate $\varepsilon_P$ at time $k+1$ according to \eqref{56}. If $\varepsilon_P(k+1) >\zeta_{CRLB}^P $, enable the variable formation enhanced obstacle position estimation (VFEO) algorithm to find the expected position of each UAV in the formation that can minimize $\varepsilon_P$ at time $k+1$, and obtain the control velocity at time $k$.

In this paper, the optimal position problem of UAVs in formation for accurate sensing of the obstacle at time $k+1$ is formulated as a constrained nonlinear optimization problem, with the goal of minimizing the position estimation error of the obstacle under the motion and formation constraints. The problem can be expressed as
\begin{align} \label{58}
	\mathop{\arg\min}_{\boldsymbol{u}}&\mathscr{F}(\boldsymbol{u}(k+1))=\varepsilon_P(k+1)\\
	s.t.\nonumber &\mathscr{H}(\boldsymbol{u}(k+1)) = \|\boldsymbol{p}_l(k+1) - \boldsymbol{u}_i(k+1)\|-r_f = 0 \tag{57a}, \\
	&\mathscr{J}(\boldsymbol{u}(k+1)) =  z_l(k+1)-z_i(k+1) = 0 \tag{57b}, \\
	&\mathscr{P}(\boldsymbol{u}(k+1)) = v_{\mathrm{max}} - \|\boldsymbol{u}_i(k+1) - \boldsymbol{u}_i(k)\| \geq 0 \tag{57c},\\
	&\mathscr{G}(\boldsymbol{u}(k+1)) =  \|\boldsymbol{u}_i(k+1) - \boldsymbol{u}_j(k+1)\|-r_{\mathrm{min}} \geq 0  \tag{57d},\\
	&\mathscr{S}(\boldsymbol{u}(k+1)) = \|\boldsymbol{s}(k+1) - \boldsymbol{u}_i(k+1)\| - r_s \geq 0 \tag{57e},
\end{align}
where $i,j=1,2,\ldots,P$ and $i\neq j$, $\boldsymbol{p}_l$ is the position of the virtual-leader. (57a) is the UAV formation limit, and ensures that the formation flies around a virtual circle with center $\boldsymbol{p}_l$ and radius $r_f$. (57b) ensures that the UAV formation is always at the same height as the virtual-leader during the formation transform. (57c) is the maximum flight speed $v_{\mathrm{max}}$ limit of UAVs. (57d) and (57e) are the minimum distance constraints between UAVs in the formation and between UAVs and the obstacle, respectively.

Using the exterior penalty function method to solve optimization problems. Construct an auxiliary function to convert the constraints of the original problem into an unconstrained problem of minimizing the auxiliary function. Define the auxiliary function as follows.
\begin{align} \label{59}
	\mathscr{Q}(\boldsymbol{u}(k+1))=&\mathscr{F}(\boldsymbol{u}(k+1))+\mu \sum\limits_{h=1}\limits^{P} \mathscr{H}_h^2(\boldsymbol{u}(k+1))\nonumber\\
	&+\mu \sum\limits_{j=1}\limits^{P} \mathscr{J}_j^2(\boldsymbol{u}(k+1)) \nonumber\\
	&+\mu \sum\limits_{p=1}\limits^{P} (\max\{0, -\mathscr{P}_p(\boldsymbol{u}(k+1))\})^2 \nonumber\\
	&+\mu \sum\limits_{g=1}\limits^{C_P^2} (\max\{0, -\mathscr{G}_g(\boldsymbol{u}(k+1))\})^2 \nonumber\\
	&+\mu \sum\limits_{s=1}\limits^{P} (\max\{0, -\mathscr{S}_s(\boldsymbol{u}(k+1))\})^2,
\end{align}
where $\mu$ is a large number, $C_P^2$ is a combination. The optimization problem is then converted to solving unconstrained $\min{\mathscr{Q}(\boldsymbol{u}(k+1))}$, which can be solved using the steepest descent method \cite{48zhang2011solving}.

\begin{figure}[!t]
	\centering
	\includegraphics[width=4in]{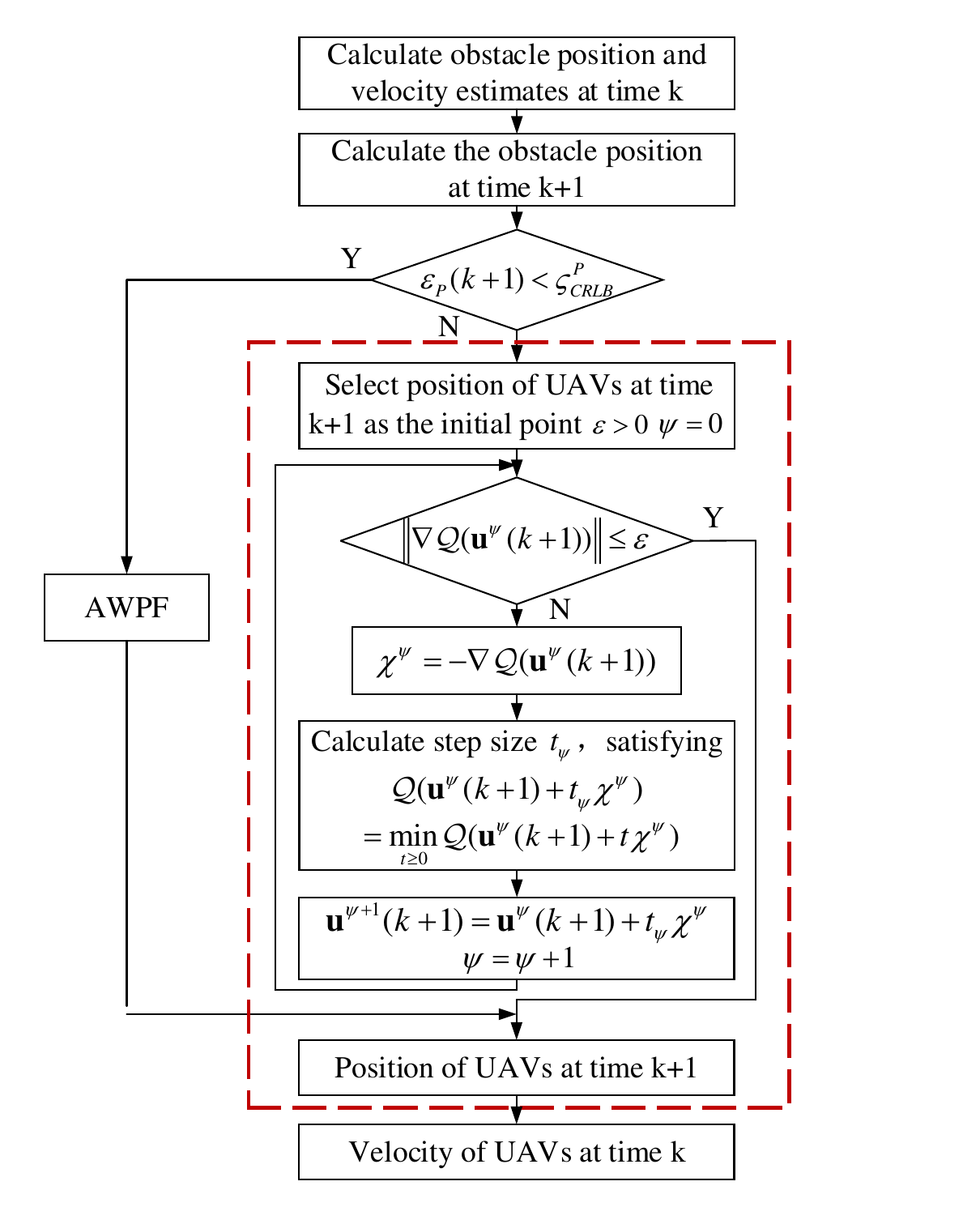}
	\caption{VFEO algorithm flowchart.}
	\label{fig_4}
\end{figure}

The flowchart of VFEO algorithm is shown in Fig. \ref{fig_4}, and the unconstrained problem $\min{\mathscr{Q}(\boldsymbol{u}(k+1))}$ in the red dashed box is solved by the steepest descent method. Where $\varepsilon$ is the termination error, $\psi$ is the number of iterations. $\chi^\psi$ is the search direction, and $t_\psi$ is the step length along the search direction.

\subsection{Complexity Analysis of the VFEO Algorithm}

The VFEO algorithm primarily consists four steps, as depicted in Fig. \ref{fig_4}. 1) Estimating obstacle distance and radial velocity. 2) Computing obstacle position and velocity. 3) Calculating the CRLB. 4) Obtaining optimal UAV positioning.

Each UAV performs $M_j$ times $N_j$-point IFFT and $N_j$ times $M_j$-point FFT on the received ISAC signals to estimate the range and radial velocity of the obstacle. This results in a total complexity of $\mathcal{O}\left(P(M_j \times N_j \times \log(M_j) \times \log(N_j))\right)$ for 2D FFT performed by $P$ UAVs in the formation. The complexity of MUAV computing the position and velocity of the obstacle primarily arises from the TWLS algorithm. Assuming the target dimension is $D$, with a total of $K$ iterations, the time complexity of the initial estimation is $\mathcal{O}(P \times D \times K \times \kappa)$, and the complexity of the weighted least squares iterations is $\mathcal{O}(P \times D \times K \times \kappa^2)$. According to \eqref{48}$-$\eqref{55}, the complexity of calculating the CRLB is  $\mathcal{O}(D^3 \times \kappa^3 +DP^2 \times \kappa^2)$. When obstacle sensing is inaccurate, the position of VFEO UAVs are obtained by solving the unconstrained problem \eqref{59}. The complexity of the steepest descent can be expressed as $\mathcal{O}\left((D \times \kappa^2 + (D+P) \times \kappa)\varepsilon ^{-2}\right)$, where $\varepsilon$ is the given accuracy. Thus, the complexity of the VFEO algorithm is $\mathcal{O}\left((D^3 \varepsilon^{-2} )\kappa^3 + (P^2\varepsilon^{-2}+PK)D\kappa^2+PMN\right)$.

It is evident that the complexity of the VFEO algorithm is mainly affected by the target dimension and accuracy. However, the problems in this paper are studied in 3D space with $D=3$. Therefore, the complexity of the VFEO algorithm depends on the accuracy given by the gradient descent algorithm.

\subsection{Online Obstacle Avoidance}
When the UAV formation performs the path-following task, an obstacle appears randomly in the space. Designing the obstacle avoidance process as the reward function is a sparse reward problem, which is inefficient for directly joining the DRL path-following model for training, and it can easily cause obstacle avoidance failure. Therefore, this paper considers the obstacle avoidance subtask an online process without pretraining.

Specifically, obstacle avoidance is triggered when the distance between the obstacle and the UAV $i$ in the formation is less than the minimum safety distance $r_{s}$. The obstacle avoidance control velocity $\boldsymbol{v}_1$ is given by \cite{49santos2017novel}
\begin{equation}\label{61}
	\begin{aligned}  
		\boldsymbol{v}_1=\boldsymbol{J}_1(\boldsymbol{u}_i)^\dagger \lambda_1(r_{s}-r_i) = \frac{\boldsymbol{u}_i -\boldsymbol{s}}{r_i} \lambda_1(r_{s}-r_i) ,
	\end{aligned}
\end{equation}
where $\lambda_1$ is the obstacle avoidance subtask gain.  $\boldsymbol{J}_1(\boldsymbol{u})$ is the Jacobian matrix, and we have
\begin{equation}\label{62}
	\begin{aligned}  
		\boldsymbol{J}_1(\boldsymbol{u}_i)^\dagger = \boldsymbol{J}_1(\boldsymbol{u}_i)^\mathrm{T}\big[\boldsymbol{J}_1(\boldsymbol{u}_i)\boldsymbol{J}_1(\boldsymbol{u_i})^\mathrm{T}\big]^{-1} = \boldsymbol{J}_1(\boldsymbol{u}_i)^\mathrm{T} .
	\end{aligned}
\end{equation}

\section{Proposed Hierarchical Subtasks Fusion Strategy}

In this section, we propose a null space based (NSB) hierarchical subtasks fusion (N-HSF) strategy. The core of the NSB method is to project low-priority subtasks to the null space of high-priority subtasks and use the fused subtasks output as the final task output, offering real-time performance and no conflict between subtasks \cite{49santos2017novel}.

Set the priority of subtasks as obstacle avoidance $=$ obstacle sensing $>$ path-following. NSB method can ensure partial completion of the path-following subtask while completing the obstacle sensing and avoidance subtasks. Obstacle avoidance is the primary security guarantee for the UAV formation to perform path-following subtask, and variable formation can enhance the position estimation of the obstacle, which can greatly improve the accuracy of obstacle sensing and ensure the success of obstacle avoidance subtask. Therefore, obstacle sensing and avoidance subtasks both belong to the highest priority.

\begin{figure}[t]
	\centering
	\includegraphics[width=3.4in]{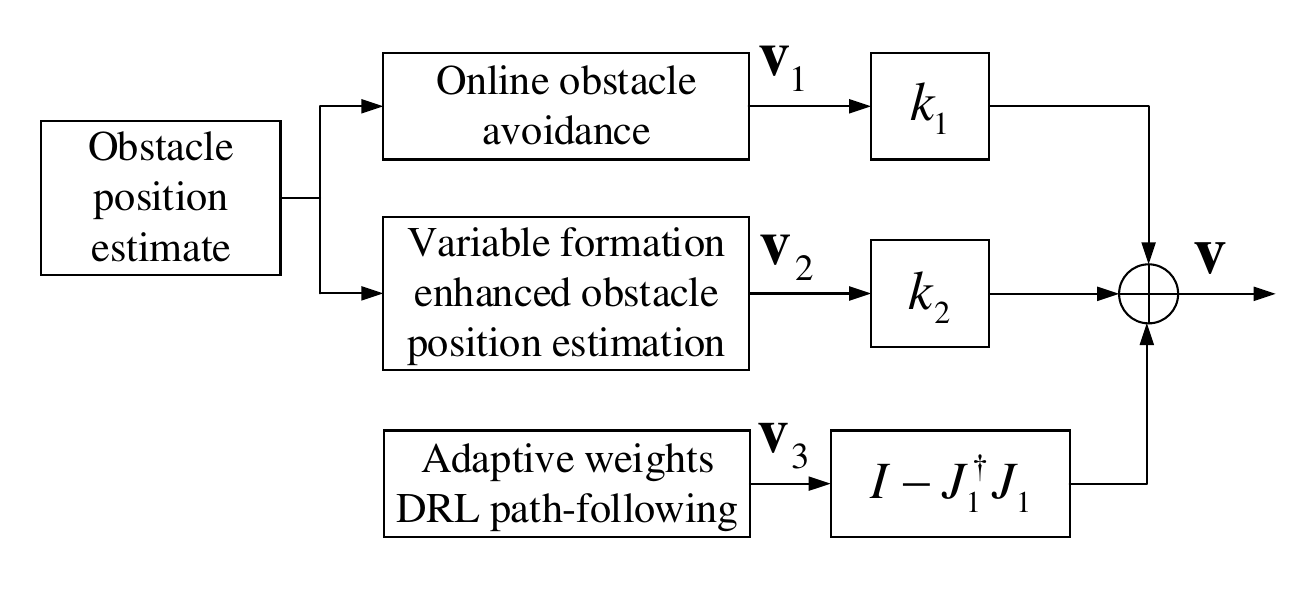}
	\caption{Hierarchical subtasks fusion strategy.}
	\label{fig_5}
\end{figure}

The proposed N-HSF strategy is as follows.
\begin{equation}\label{63}
	\begin{aligned}  
		\boldsymbol{v}=k_1\boldsymbol{v}_1+k_2\boldsymbol{v}_2+(\mathbf{I}-\boldsymbol{J}_1^\dagger\boldsymbol{J}_1)\boldsymbol{v}_3 ,
	\end{aligned}
\end{equation}
where $k_1$ and $k_2$ are the weights of $\boldsymbol{v}_1$ and $\boldsymbol{v}_2$, respectively. $\boldsymbol{v}_1$ is given by \eqref{61} (Section IV-E), $\boldsymbol{v}_2$ is given by the VFEO algorithm (Section IV-C), and $\boldsymbol{v}_3$ is given by the AWPF algorithm (Section IV-A). The N-HSF strategy is shown in Fig. \ref{fig_5}.

Since sensing and obstacle avoidance both belong to the highest priority, the two subtasks are not projected into the null space of each other. Path-following subtask has the lowest priority, and it is considered to be projected into the null space of obstacle sensing and avoidance. However, performing the sensing subtask does not affect the path-following of each UAV in the formation. Because MUAV assigns the VFTs of UAVs, the sensing subtask can be completed by changing the VFTs. Therefore, it is only necessary to project path-following into the null space of the obstacle avoidance subtask.

In \eqref{63}, $\boldsymbol{v}$ is the final hierarchical fusion output of the three subtasks, which is the control velocity of UAVs at the next time. Subtract the current velocity from the next control velocity to obtain the current control acceleration $\boldsymbol{\alpha}_i(k)=[a_{x,i},a_{y,i},a_{z,i}]_k^\mathrm{T}$, which is the virtual control variable derived in the UAV Kinetics model.

\section{Simulation Results and Analysis}

In this section, we provide the simulation results of the AWPF algorithm  and conduct a comparison with reference algorithms, along with an analysis of their complexities. Then, the performance of the VFEO algorithm and online obstacle avoidance algorithm are given on the basis of DRL path-following. Meanwhile, we also analyze the feasibility of the N-HSF strategy.

We consider a formation consisting of five UAVs, with one MUAV and four AUAVs. The OFDM waveforms modulated by DM-RS sequences used by UAVs support dmrs-AdditionalPosition = pos3 in \cite{50gpp2018nr}. The simulation parameters are shown in Table 1.
\begin{table}[!h]
	\renewcommand{\arraystretch}{1}
	\caption{Simulation Parameters}
	\label{table_1}
	\centering
	\setlength{\tabcolsep}{5mm}{
		\begin{tabular}{l l}
			\toprule
			\textbf{\textbf{Parameters}} & \textbf{Value} \\
			\midrule
			\textbf{UAV formation parameters} &  \\
			Formation radius $r_f$ & 20 m \\
			Minimum formation safety distance $r_{\mathrm{min}}$ & 5 m \\
			Minimum obstacle avoidance safety distance $r_s$ & 5 m \\
			Positioning accuracy threshold $\scriptstyle \zeta_{CRLB}^P$ & 0.5 m \\
			\specialrule{0em}{2pt}{2pt}
			
			\textbf{ISAC signal parameters} &  \\
			SNR & 20 dB \\
			Carrier frequency $f_c$ & 24 GHz \\
			Frequency spacing of subcarriers $\scriptstyle{\triangle}$$f$ & 120 kHz \\
			Total duration of OFDM symbol $T_s$ & 8.92 $\mu$s \\
			Number of subcarriers $N$ & 256 \\
			Number of OFDM symbols $M$ & 140 \\
            Number of subcarriers carrying DM-RS $N_J$ & 128 \\
			Number of OFDM symbols carrying DM-RS $M_J$ & 40 \\
			\specialrule{0em}{2pt}{2pt}
			
			\textbf{DDPG parameters} &  \\
			Learning rate for actor &  1e-4 \\
			Learning rate for critic & 1e-3 \\
			Discount factor & 0.99 \\
			Batch size  & 128 \\
			Soft replacement & 0.01 \\
			Memory capacity & 50000 \\
			\bottomrule
	\end{tabular}}
\end{table}

\subsection{UAV Formation Path-Following}

The simulation time is 400 s, the control cycle is 1 s, and the control speed of UAV $i$ is $|\dot{\boldsymbol{u}}_i| \in [0,78]$ m/s.

The position initialization of the virtual-leader is $\!\boldsymbol{p}_l(0)=[100,100,50]^\mathrm{T}\!$, and the position initialization of the five UAVs are $\!\boldsymbol{u}_1(0)=[0,20,0]^\mathrm{T}\!$, $\!\boldsymbol{u}_2(0)=[20,0,0]^\mathrm{T}\!$, $\!\boldsymbol{u}_3(0)=[10,10,0]^\mathrm{T}\!$, $\!\boldsymbol{u}_4(0)=[15,5,0]^\mathrm{T}\!$, and $\!\boldsymbol{u}_5(0)=[5,15,0]^\mathrm{T}\!$, respectively, where $\!\boldsymbol{u}_1$ represents the position of MUAV. Moreover, the velocity initialization of all UAVs in the formation are $\!\dot{\boldsymbol{u}}(0)=[0,0,0]^\mathrm{T}\!$. 
\begin{figure}[!t]
	\centering
	\includegraphics[width=3.2in]{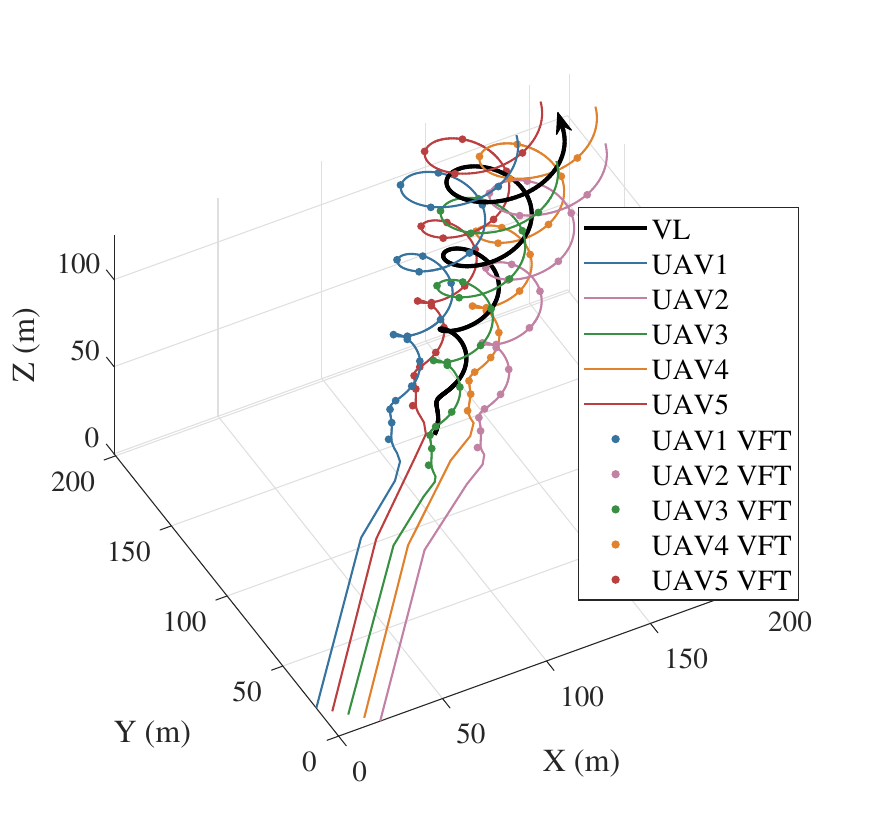}
	\caption{Path-following trajectory using AWPF algorithm.}
	\label{fig_6}
\end{figure}

\begin{figure}[!t]
	\centering
	\includegraphics[width=3.15in]{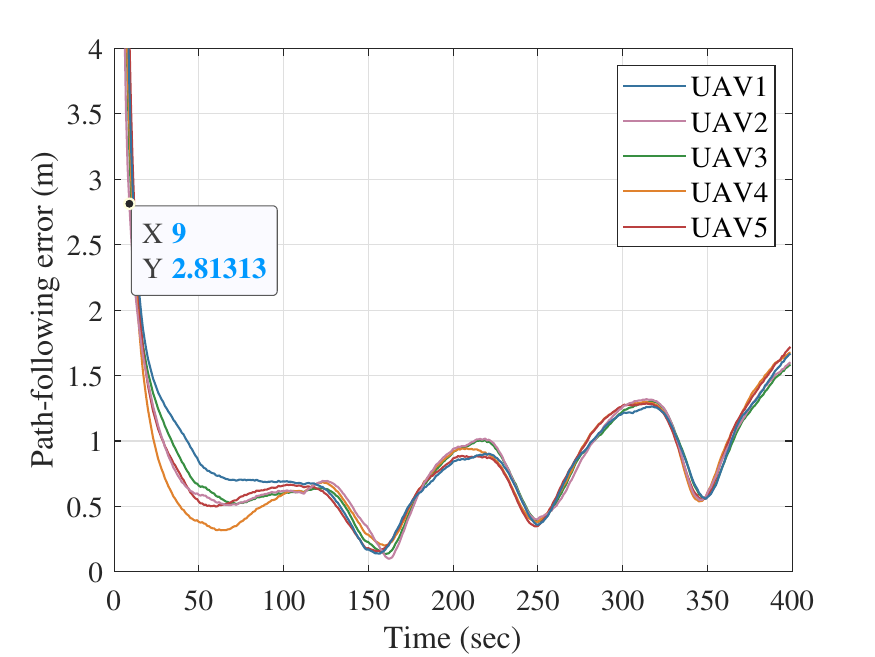}
	\caption{Path-following error using AWPF algorithm.}
	\label{fig_7}
\end{figure}

The path-following subtask comprises two phases, chasing and following. Fig. \ref{fig_6} illustrates the path-following trajectory of the UAV formation during the simulation period, which includes the actual trajectories provided by the AWPF algorithm, the VFT trajectories of UAVs assigned by MUAV, and the virtual-leader's trajectory. Fig. \ref{fig_7} presents the following error of the actual trajectory. We posit that once the path-following error drops below 3 m, the UAV formation transitions from the chasing phase to the following phase. In 0$-$9 s, the UAV formation is in the chasing phase and gradually approaches the virtual-leader. In 10$-$400 s, the UAVs assume a predetermined formation and follow the virtual-leader. Fig. \ref{fig_7} indicates that the following errors of all UAVs in the formation converge to a bounded range of less than 1.7 m and remain consistent.
\begin{figure}[!t]
	\centering
	\includegraphics[width=3.15in]{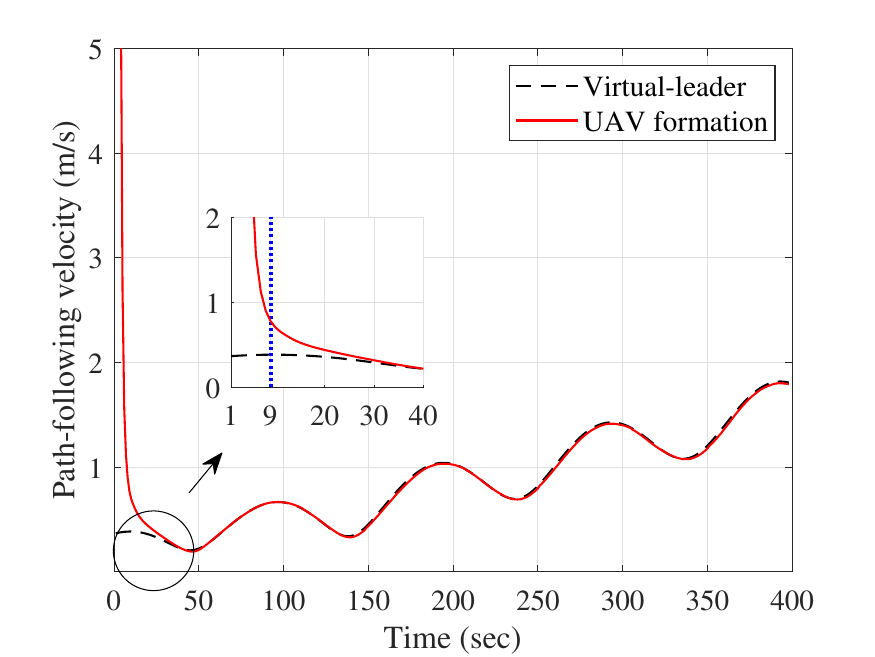}
	\caption{Path-following velocity using AWPF algorithm.}
	\label{fig_v}
\end{figure}

Fig. \ref{fig_v} shows the average velocity of the UAV formation during the AWPF. At $k = 1$ s, the reward function is dominated by the distance errors between the UAVs and the VFTs, with the minimum weight $\omega_2$ assigned to the velocity error $e_{v,ie}$. At this time, the UAV formation chases the virtual-leader at maximum velocity. As $e_{d,ie}$ decreases, $\omega_2$ increases gradually, causing the UAVs to decelerate in preparation for accurate following. The velocity of the UAV formation from rapid descent to a smoother pattern at 9 s, indicating the end of the chasing phase and the commencement of the following phase. In this instance, $\omega_2$ reaches its maximum. The reward function weights for the AWPF algorithm are
\begin{equation}\label{64}
	{\omega_1} = \begin{cases}
		0.05,&{\text{if}}\ e_{d,ie}\leq3, \\ 
		e_{d,ie}/(e_{d,ie}+40),&{\text{otherwise}} ,
	\end{cases}
\end{equation}
\begin{equation}\label{65}
	{\omega_2} = \begin{cases}
		0.95,&{\text{if}}\ e_{d,ie}\leq3, \\ 
		40/(e_{d,ie}+40),&{\text{otherwise.}} 
	\end{cases}
\end{equation}

After 35 s, the velocities of the UAVs and the virtual-leader converge, indicating that the formation accurately follows the virtual-leader and mimics its velocity.

\begin{figure}[!t]
	\centering
	\includegraphics[width=3.15in]{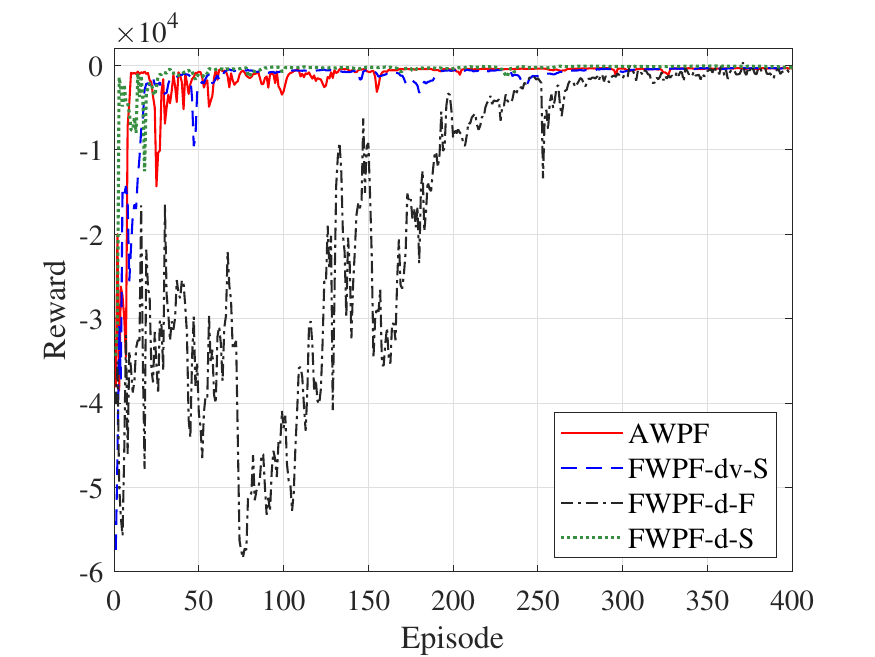}
	\caption{Reward.}
	\label{fig_8}
\end{figure}

\begin{figure}[!t]
	\centering
	\includegraphics[width=3.15in]{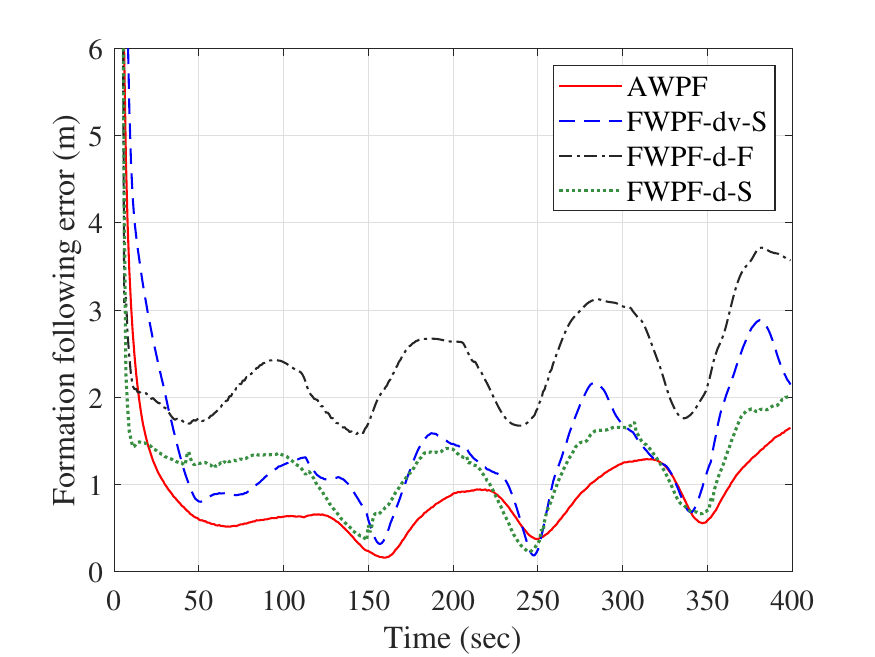}
	\caption{Average path-following error of UAVs.}
	\label{fig_9}
\end{figure}

To validate the superiority of the proposed AWPF algorithm, it is compared against three reference algorithms. In Fig. \ref{fig_8} and \ref{fig_9}, FWPF represents path-following with a fixed weight reward function, -d signifies the sole consideration of distance error in the reward function, while -dv denotes both distance and velocity errors are considered. -S and -F indicate training for an individual UAV and for the UAV formation respectively. The reference algorithms also consist of five UAVs and employ identical simulation parameters as AWPF. In the FWPF-dv-S algorithm, the weights are fixed as $\omega_1=0.05$ and $\omega_2=0.95$. We validate the superiority of AWPF in terms of convergence speed and path-following error.

Fig. \ref{fig_8} illustrates the comparison of rewards obtained by the AWPF and the reference algorithms in each episode. It can be seen that after a period of training, the FWPF-d-S algorithm first converges at 85 episodes, attributed to the consideration of only distance error $e_{d,ie}$ in the reward function. Subsequently, the AWPF and FWPF-dv-S algorithms converge at 160 and 265 episodes respectively. These two algorithms consider both distance and velocity errors, with the distinction lying solely in the adaptive adjustment of reward function weights. The FWPF-d-F algorithm exhibits the slowest convergence speed. This is attributed to its reliance on formation training, which requires ensuring that UAVs form a specific formation to perform tasks. In contrast, the training algorithm for individual UAV aims to obtain a formation-generic model, with the allocation of VFTs handled by MUAV.

Fig. \ref{fig_9} compares the average path-following errors between the AWPF and the reference algorithms. The AWPF algorithm achieves the minimal average following error for the UAV formation, converging within 1.7 m. The following errors for the FWPF-d-S and FWPF-dv-S algorithms are at a moderate level, measuring 2 m and 2.8 m respectively. In contrast, the FWPF-d-F algorithm, due to the necessity of adhering to formation requirements, exhibits the worst following performance. This demonstrates that the use of adaptive weights in the reward function is crucial for achieving effective path-following performance. On the one hand, the FWPF-d-S and FWPF-d-F (FWPF-d-S$\&$F) algorithms can chase the virtual-leader more quickly because they maintain maximum speed during the chasing phase. In contrast, the chasing speeds of the AWPF and FWPF-dv-S algorithms are slower due to the partial influence of $e_{v,ie}$ in the reward function, despite $e_{d,ie}$ predominating. AWPF adaptively maintains a smaller $\omega_2$ during the chasing phase as demonstrated by \eqref{65}, bringing its chasing speed closer to the FWPF-d-S$\&$F algorithms. Conversely, the FWPF-dv-S algorithm, with the fixed weights, experiences a significant reduction in chasing speed. On the other hand, once the UAV formation catches up with the virtual-leader and enters the following phase, the AWPF algorithm has a larger $\omega_2$ than the chasing phase. The UAV formation minimizes both $e_{d,ie}$ and $e_{v,ie}$ to achieve reduced following errors. Numerical analyses reveal that compared to the reference algorithms, the AWPF algorithm enhances following accuracy by 21\% to 124\% while achieving rapid convergence.

Next, we derive and analyze the complexity of the reference algorithms for AWPF. The reference algorithms share the same network architecture as AWPF and possess an equivalent neural network complexity. The FWPF-d-F algorithm does not involve the process of VFTs calculation but requires computing the desired formation, which has the complexity of $\Psi_{\texttt{f}}=\Psi_{\texttt{VFT}}$. Similar to AWPF, FWPF-dv-S requires considering both distance and velocity errors. Therefore, the computational complexity of FWPF-dv-S is the same as AWPF. In contrast, FWPF-d-S$\&$F algorithms exclude velocity error calculation, and the complexities are reduced by one term of $\Psi_\texttt{e}$ compared to AWPF, as shown in \eqref{24}.
	\begin{equation}\label{24}
		\Psi_\texttt{Fd} = \Psi_{\texttt{VFT}}+\Psi_{\texttt{e}}+\Psi_{\texttt{DRL}}
		\approx P(\sum_{l=1}^{L_{n}}S_{l-1} \times S_{l})\kappa^2+9P\kappa.
	\end{equation}
It can be inferred that the complexities of the four algorithms are approximately equal based on \eqref{23} and \eqref{24}. AWPF enhances path-following performance without increasing algorithmic complexity.


\subsection{Variable Formation Enhanced Obstacle Position Estimates}
\begin{figure}[!t]
	\centering
	\includegraphics[width=3.15in]{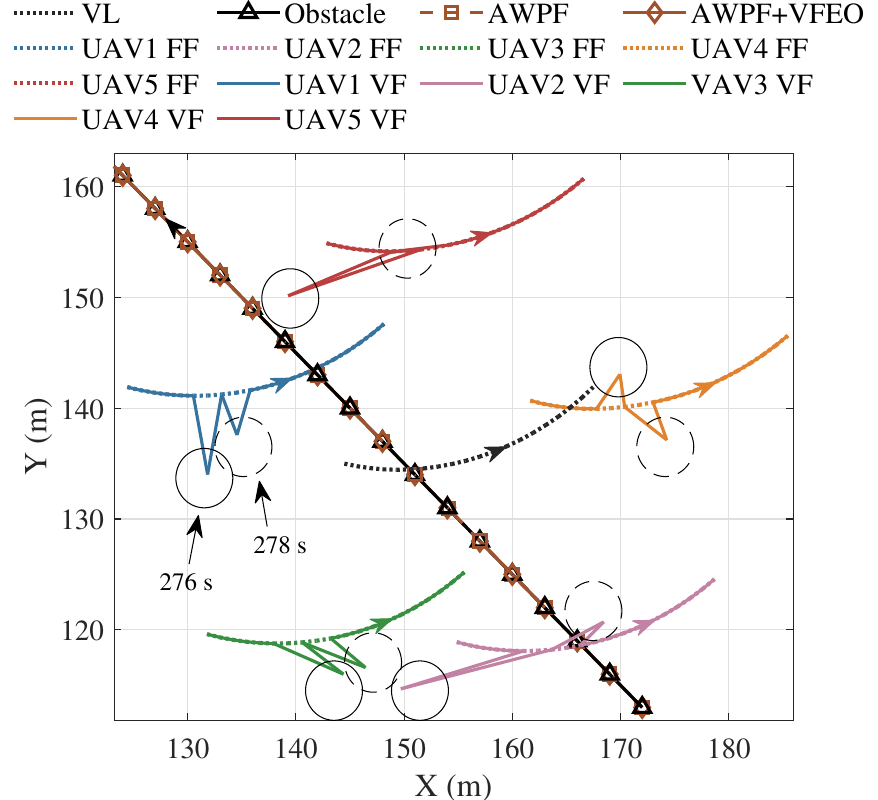}
	\caption{X-Y axis view of variable formation enhanced obstacle position estimation.}
	\label{fig_10}
\end{figure}
\begin{figure}[!t]
	\centering
	\includegraphics[width=3.15in]{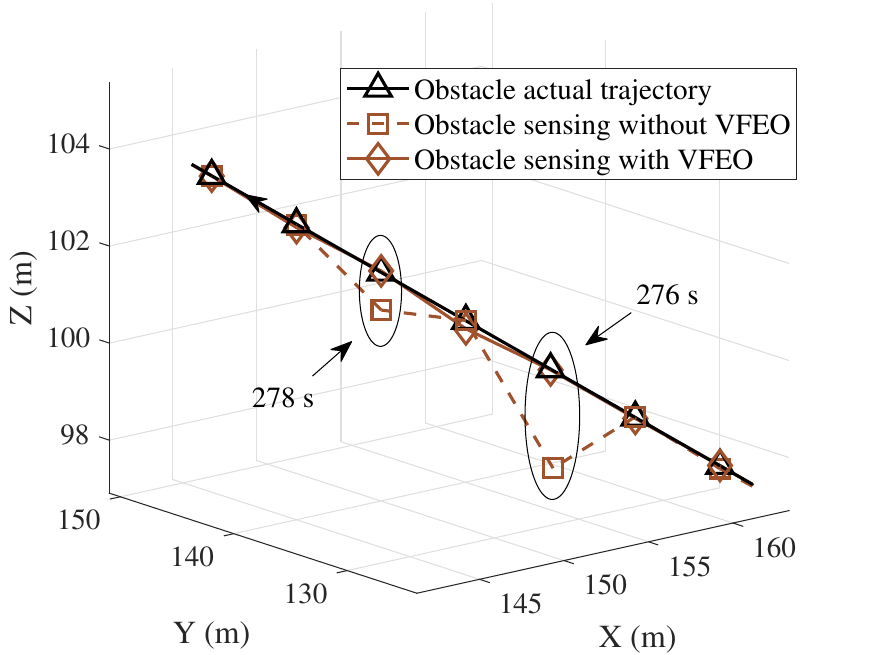}
	\caption{VFEO position estimation in the time range of 274$-$280 s.}
	\label{fig_18}
\end{figure}
\begin{figure}[!t]
	\centering
	\includegraphics[width=3.15in]{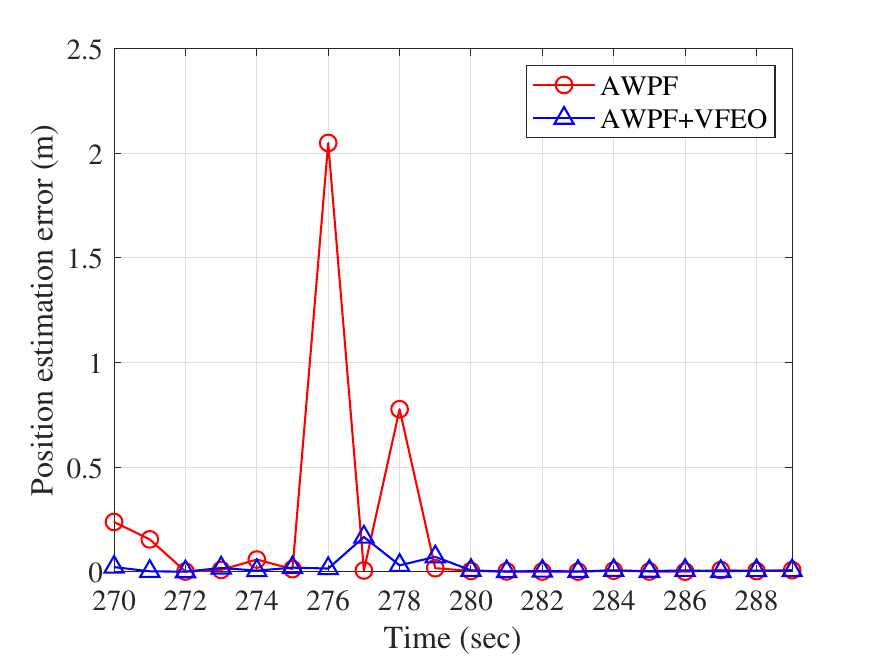}
	\caption{Obstacle positioning error.}
	\label{fig_11}
\end{figure}
\begin{figure}[!t]
	\centering
	\subfloat[276 s]{
		\includegraphics[width=2.9in]{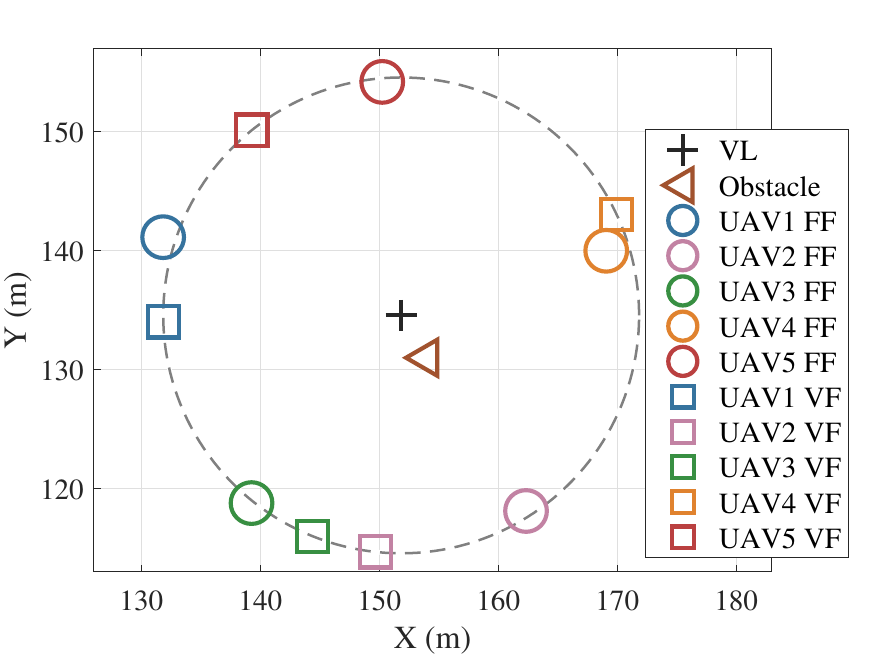}} \\
	\subfloat[278 s]{
		\includegraphics[width=2.9in]{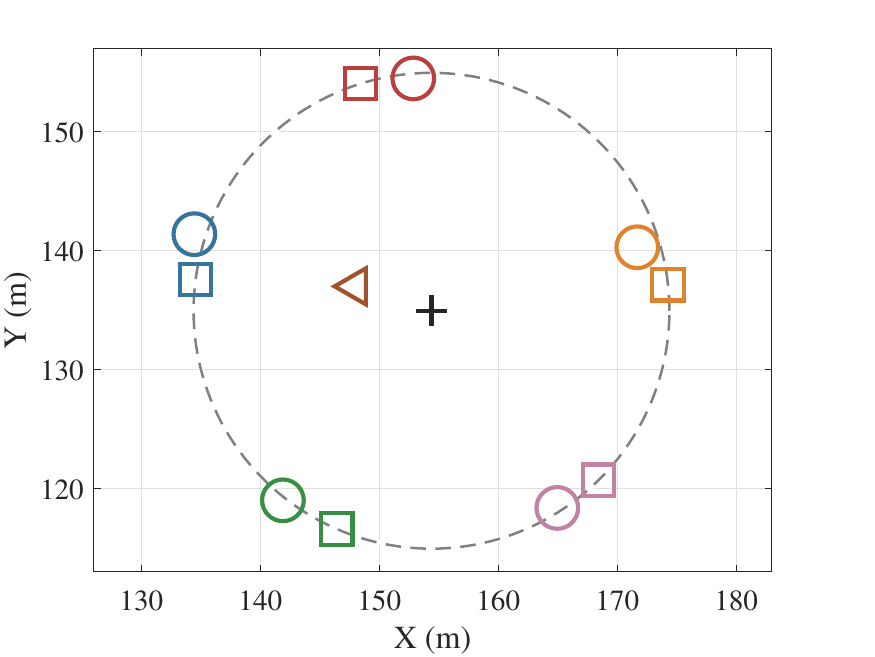}} 
	\caption{Position of UAVs at 276 s and 278 s.}
	\label{fig_13}
\end{figure}
To validate the performance of the VFEO algorithm in cases of inaccurate obstacle sensing, we transform the UAV formation based on AWPF path-following trajectories to enhance obstacle sensing capabilities. At 270 s of path-following, a spatial obstacle appears. The obstacle position is initialized as $\!\boldsymbol{s}_1(270)=[172,113,94]^\mathrm{T}\!$, with a velocity of $\!\dot{\boldsymbol{s}}_1(270)=[-3,3,1]^\mathrm{T}\!$, and the observation time of 20 s. The actual trajectory of the obstacle is shown as the black triangular line in Fig. \ref{fig_10} and Fig. \ref{fig_18}, and the obstacle sensing by fixed-formation (FF) and variable-formation (VF) are shown as the square dashed line and the diamond-shaped solid line respectively. The unmarked dotted line indicates the FF path-following trajectories of the UAV formation using the AWPF algorithm, and the unmarked solid line indicates the joint VF trajectories of the AWPF and VFEO algorithms. The arrows solely indicate the direction of motion. Fig. \ref{fig_18} is a section taken from Fig. \ref{fig_10}, and it can be clearly seen that the accuracy of obstacle sensing by variable formation is substantially improved at 276 s and 278 s.
\begin{figure}[!t]
	\centering
	\includegraphics[width=3.1in]{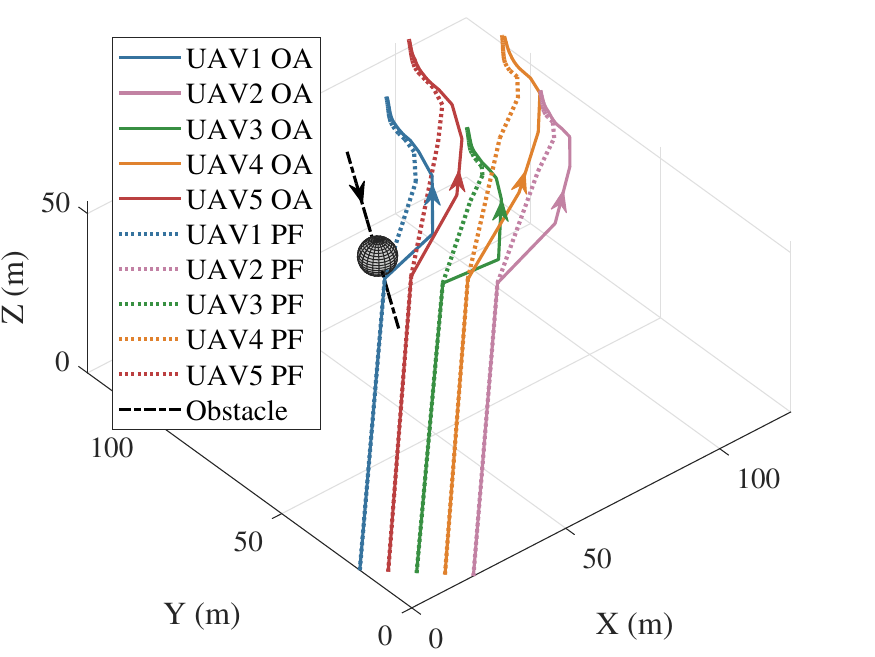}
	\caption{Obstacle avoidance trajectory in chasing phase.}
	\label{fig_14}
\end{figure}
\begin{figure}[!t]
	\centering
	\includegraphics[width=3.1in]{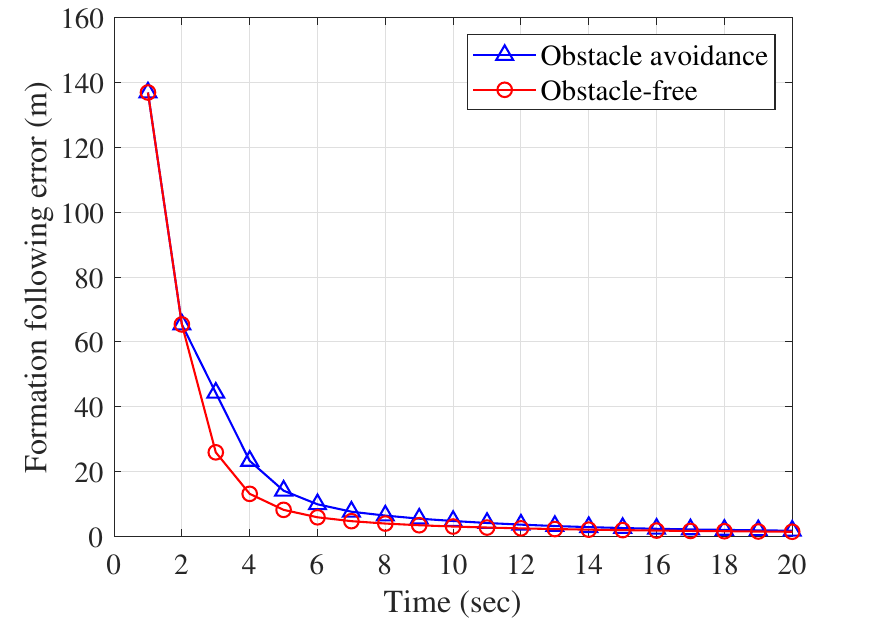}
	\caption{Formation following error of obstacle avoidance in chasing phase.}
	\label{fig_15}
\end{figure}

As shown by the red line with circular markers in Fig. \ref{fig_11}, at 276 s and 278 s, the estimation errors of the FF on the obstacle position are relatively large, which are 2.05 m and 0.78 m. Excessive sensing error can adversely affect obstacle avoidance performance and even lead to obstacle avoidance failure. At 275 s and 277 s, the VF uses the VFEO algorithm to calculate that the position estimation errors at next time, which gives $\varepsilon_P(276)>\zeta_{CRLB}^P$ and $\varepsilon_P(278)>\zeta_{CRLB}^P$, and perform the transformation of UAV formation. The VFEO algorithm finds the optimal position of UAVs in the formation at next time, thereby enhancing the performance of UAV formation in sensing obstacles. As shown in Fig. \ref{fig_13}, (a) and (b) represent the distribution of UAVs in the formation at 276 s and 278 s respectively. When the CRLB does not satisfy the sensing performance requirement, the UAV formation no longer senses the obstacle in a uniformly distributed FF (circles arranged along the dashed line), but finds the optimal distribution of UAVs that enables more accurate obstacle observation by the VFEO algorithm, as shown by squares distributed on the dashed circle in Fig. \ref{fig_13}. By transforming the UAV formation, the obstacle position sensing errors at 276 s and 278 s are reduced to 0.02 m and 0.03 m, respectively, as shown by the blue triangular line in Fig. \ref{fig_11}, which significantly improved the obstacle sensing performance.

We can see in Fig. \ref{fig_10} that the VFEO algorithm is performed twice on the basis of AWPF path-following trajectories at 276 and 278 s. After the VFEO algorithm completes, the UAV formation returns to the following path of the AWPF algorithm in 277 s and 279$-$289 s. There is no conflict between the path-following and obstacle sensing subtasks.

\subsection{Obstacle Avoidance and Hierarchical Subtasks Fusion}
\begin{figure}[!t]
	\centering
	\includegraphics[width=3.1in]{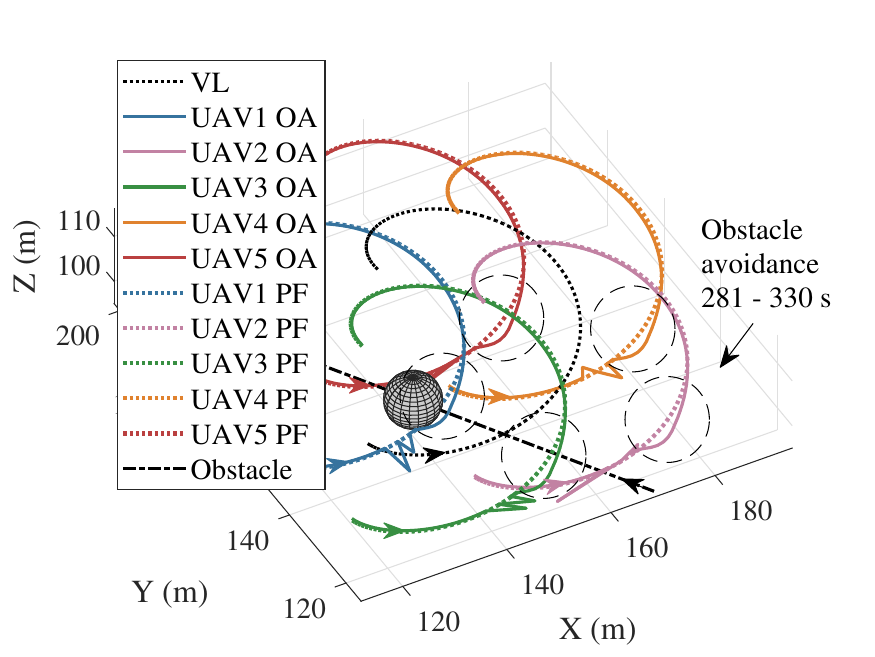}
	\caption{Obstacle avoidance trajectory in following phase.}
	\label{fig_16}
\end{figure}
\begin{figure}[!t]
	\centering
	\includegraphics[width=3.1in]{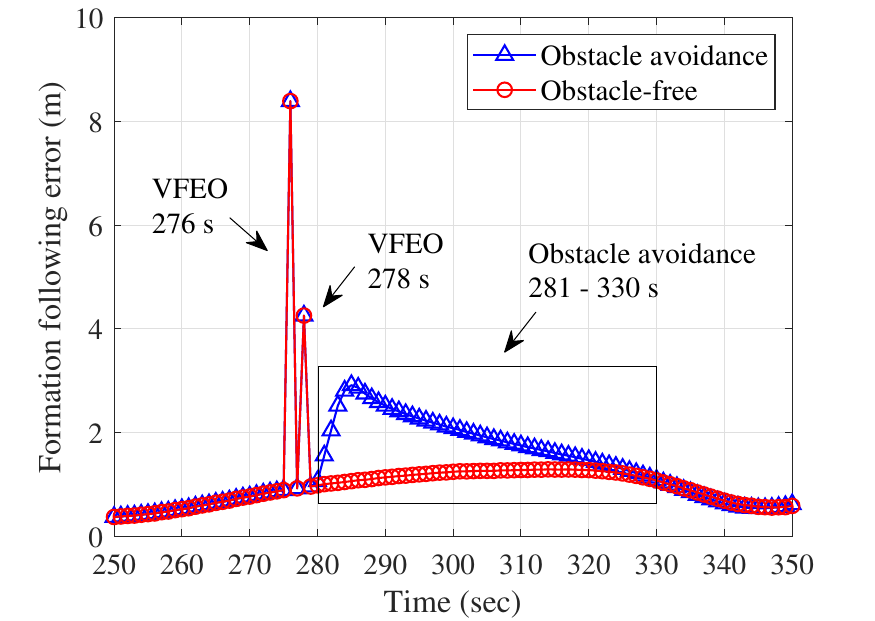}
	\caption{Formation following error of obstacle avoidance in following phase.}
	\label{fig_17}
\end{figure}
In this subsection, on the basis of DRL path-following and variable formation enhanced obstacle position estimation, the performance of online obstacle avoidance is analyzed, and the availability of the hierarchical subtasks fusion strategy is verified. In the virtual-leader path-following process in Fig. \ref{fig_6}, two dynamic obstacles are added, including $\boldsymbol{s}_2=[90,160,70]^\mathrm{T}$ appears in 1$-$6 s for the chasing phase, with the velocity of $\dot{\boldsymbol{s}}_2=[-10,-20,-10]^\mathrm{T}$, and $\boldsymbol{s}_1$ (the detailed description can be found in Section VI-B), which appears in the following phase. The gray spheres in Fig. \ref{fig_14} and Fig. \ref{fig_16} represent $\boldsymbol{s}_2$ and $\boldsymbol{s}_1$, respectively. The dash-dotted line is the trajectory of the dynamic obstacle, the dashed lines are the AWPF path-following trajectories when the obstacle avoidance subtask is not triggered, and the solid lines are the actual obstacle avoidance trajectories based on AWPF and VFEO algorithms. In Fig. \ref{fig_16}, we place significant emphasis on the obstacle avoidance trajectories enclosed by the dashed circles.

As shown in Fig. \ref{fig_15}, at 2 s, the distance between $\boldsymbol{s}_2$ and the UAV formation is less than the safety distance $r_{s}$, triggering obstacle avoidance. After the obstacle avoidance process lasted 15 s, it returned to the DRL path-following phase at 16 s. As shown in Fig. \ref{fig_17}, in 250$-$280 s, the obstacle is outside the obstacle avoidance range of the formation, and each UAV carries out path-following according to the AWPF model. The abrupt increases in path-following error at 276 s and 278 s are attributed to performing the VFEO algorithm, which enhance the sensing of obstacle $\boldsymbol{s}_1$ (in conjunction with Fig. \ref{fig_10}). In 281$-$330 s, the obstacle $\boldsymbol{s}_1$ enters the emergency range, and UAV formation performs the obstacle avoidance subtask. During this period, the following error increased to a maximum of 2.92 m and then gradually decreased. After 330 s, UAV formation enters the normal path-following phase. In the process of obstacle avoidance, UAVs keep a safety distance from obstacles and do not collide.

It is worth noting that in the path-following process after obstacle avoidance, the UAV formation following error is consistent with the original AWPF path-following error, as shown in Fig. \ref{fig_15} at 16$-$20 s and Fig. \ref{fig_17} at 331$-$350 s. It can be seen that the obstacle avoidance subtask does not interfere with the path-following subtask. Moreover, Fig. \ref{fig_17} indicates that the transformation of UAV formation at 276 s and 278 s do not adversely affect the subsequent obstacle avoidance subtask. And we have obtained the conclusion that there is no conflict between the path-following and obstacle sensing subtasks in the previous subsection. Therefore, the proposed N-HSF possesses the ability to realize conflict-free scheduling of the three subtasks.

\section{Conclusion}

In this paper, we study a hierarchical subtasks processing for UAV formation cooperative path-following in 3D space, focusing on three subtasks of path-following, obstacle sensing, and obstacle avoidance and their hierarchical fusion scheduling. We propose the AWPF algorithm to complete the path-following subtask. A reward function with adaptive weights is designed to provide high path-following accuracy without increasing algorithm complexity or compromising convergence speed. We derive the CRLB for obstacle sensing based on ISAC signals, and propose the VFEO algorithm, which enhances obstacle position estimation by transforming the UAV formation. The online obstacle avoidance algorithm is proposed on the basis of DRL path-following to solve the sparse reward problem. In addition, the N-HSF strategy is proposed to fuse the control outputs of these subtasks. Simulation results indicate that the AWPF algorithm improves the path-following accuracy by 21$-$124\% compared to the reference algorithms. The VFEO algorithm can substantially improve the obstacle positioning to the decimeter level. The online obstacle avoidance makes it collision-free. And the N-HSF strategy ensures no conflict of subtasks. In future work, we will focus on cooperative sensing and obstacle avoidance of multiple dynamic obstacles, as well as more applications to the real world.

\bibliographystyle{IEEEtran}
\bibliography{IEEEabrv,ref}

\end{document}